\documentclass[aps,prc,twocolumn,amssymb,amsmath,amssymb,amsfonts,superscriptaddress]{revtex4-1}

\usepackage{graphicx,graphics,pstricks,epsfig,wrapfig}

\def\ra{\rightarrow}

\def\bit{\begin{itemize}}
\def\eit{\end{itemize}}
\def\beq{\begin{equation}}
\def\eeq{\end{equation}}
\def\beqn{\begin{eqnarray}}
\def\eeqn{\end{eqnarray}}
\def\nuebar{\bar\nu_e}

\begin{document}

\DeclareGraphicsExtensions{.pdf,.eps,.epsi,.jpg}

\title{Improved Predictions of Reactor Antineutrino Spectra}

\author{Th.~A.~Mueller}
\affiliation{Commissariat \`{a} l'\'{E}nergie Atomique et aux \'{E}nergies Alternatives,\\
Centre de Saclay, IRFU/SPhN, 91191 Gif-sur-Yvette, France}

\author{D.~Lhuillier}
\email{Corresponding author: david.lhuillier@cea.fr}
\affiliation{Commissariat \`{a} l'\'{E}nergie Atomique et aux \'{E}nergies Alternatives,\\
Centre de Saclay, IRFU/SPhN, 91191 Gif-sur-Yvette, France}

\author{M.~Fallot}
\affiliation{Laboratoire SUBATECH, \'Ecole des Mines de Nantes, Universit\'e de Nantes, CNRS/IN2P3, 4 rue Alfred Kastler, 44307 Nantes
Cedex 3, France}

\author{A.~Letourneau}
\affiliation{Commissariat \`{a} l'\'{E}nergie Atomique et aux \'{E}nergies Alternatives,\\
Centre de Saclay, IRFU/SPhN, 91191 Gif-sur-Yvette, France}

\author{S.~Cormon}
\affiliation{Laboratoire SUBATECH, \'Ecole des Mines de Nantes, Universit\'e de Nantes, CNRS/IN2P3, 4 rue Alfred Kastler, 44307 Nantes
Cedex 3, France}

\author{M.~Fechner}
\affiliation{Commissariat \`{a} l'\'{E}nergie Atomique et aux \'{E}nergies Alternatives,\\
Centre de Saclay, IRFU/SPP, 91191 Gif-sur-Yvette, France}

\author{L.~Giot}
\affiliation{Laboratoire SUBATECH, \'Ecole des Mines de Nantes, Universit\'e de Nantes, CNRS/IN2P3, 4 rue Alfred Kastler, 44307 Nantes
Cedex 3, France}

\author{T.~Lasserre}
\affiliation{Commissariat \`{a} l'\'{E}nergie Atomique et aux \'{E}nergies Alternatives,\\
Centre de Saclay, IRFU/SPP, 91191 Gif-sur-Yvette, France}

\author{J.~Martino}
\affiliation{Laboratoire SUBATECH, \'Ecole des Mines de Nantes, Universit\'e de Nantes, CNRS/IN2P3, 4 rue Alfred Kastler, 44307 Nantes
Cedex 3, France}

\author{G.~Mention}
\affiliation{Commissariat \`{a} l'\'{E}nergie Atomique et aux \'{E}nergies Alternatives,\\
Centre de Saclay, IRFU/SPP, 91191 Gif-sur-Yvette, France}

\author{A.~Porta}
\affiliation{Laboratoire SUBATECH, \'Ecole des Mines de Nantes, Universit\'e de Nantes, CNRS/IN2P3, 4 rue Alfred Kastler, 44307 Nantes
Cedex 3, France}

\author{F.~Yermia}
\affiliation{Laboratoire SUBATECH, \'Ecole des Mines de Nantes, Universit\'e de Nantes, CNRS/IN2P3, 4 rue Alfred Kastler, 44307 Nantes
Cedex 3, France}
\date{\today}

\begin{abstract} Precise predictions of the antineutrino spectra emitted by nuclear reactors is a key ingredient in
measurements of reactor neutrino oscillations as well as of the recent applications to the surveillance of power plants
in the context of non proliferation of nuclear weapons. We report new calculations including the latest information from
nuclear databases and a detailed error budget. The first part of this work is the so-called {\it ab initio} approach
where the total antineutrino spectrum is built from the sum of all $\beta$-branches of all fission products predicted by
an evolution code.  Systematic effects and missing information in nuclear databases lead to final relative uncertainties
in the 10 to 20\% range. A prediction of the antineutrino spectrum associated with the fission of $^{238}$U is given
based on this {\it ab initio} method. For the dominant isotopes $^{235}$U and $^{239}$Pu, we developed a more accurate
approach combining information from nuclear databases and reference electron spectra associated with the fission of
$^{235}$U, $^{239}$Pu and $^{241}$Pu, measured at ILL in the 80's. We show how the anchor point of the measured total
$\beta$-spectra can be used to suppress the uncertainty in nuclear databases while taking advantage of all the
information they contain. We provide new reference antineutrino spectra for $^{235}$U, $^{239}$Pu and $^{241}$Pu
isotopes in the 2-8 MeV range. While the shapes of the spectra and their uncertainties are comparable to that of the previous 
analysis of the ILL data, the normalization is shifted by about +3\% on average. In the perspective of the re-analysis of past 
experiments and direct use of these results by upcoming oscillation experiments, we discuss the various sources of errors 
and their correlations as well as the corrections induced by off equilibrium effects. 
\end{abstract}  

\maketitle

\section{Introduction} 
\label{sec:intro}

Nuclear power plants are the most intense man-controlled sources of neutrinos. With an average energy of about 200 MeV
released per fission and 6 neutrinos produced along the $\beta$-decay chain of the fission products, one expects some
$2\times10^{20}$ $\nu/s$ emitted in a $4\pi$ solid angle from a 1 GW reactor (thermal power). Since unstable fission
products are neutron-rich nuclei all $\beta$-decays are of $\beta^-$ type and the neutrino flux is actually pure
electronic antineutrinos ($\nuebar$). These unique features have been exploited by several neutrino oscillations
experiments~\cite{Bemporad:2001qy,Lasserre:2005}. Improvement in the accuracy of $\nuebar$ spectra is motivated by next 
generation experiments~\cite{Ardellier:2006mn,Guo:2007ug,Ahn:2010vy} aiming at unprecedented sensitivity to the last 
unknown mixing angle $\theta_{13}$. The value of this parameter may determine the future trend of the neutrino physics, in 
particular for the search of CP violation in the lepton sector. Recent developments of compact (1 m$^3$ target scale) $\nuebar$
detectors as new safeguard tools for the monitoring of reactors~\cite{Rovno,SONGS,Nucifer} would also benefit from an
accurate description of $\nuebar$ spectra.

In a reactor core, only 1 neutron among the few generated by the fission of a $^{235}$U nucleus should induce another
fission, so that the core never reaches the over-critical regime. A fraction of the neutrons is actually captured by the
dominant $^{238}$U isotope leading to the production of new fissile isotopes: $^{239}$Pu and to a lesser extent
$^{241}$Pu. When operating, a core is thus burning $^{235}$U and accumulating $^{239}$Pu. This is the so-called burnup
process. In a pressurized water reactor, fission rates from both isotopes become comparable at the end of a cycle. The
remaining fissions of $^{241}$Pu and  fast neutron induced fissions of $^{238}$U share about 10\% of the reactor power.
As a result, the accurate prediction of the $\nuebar$ spectrum of a reactor requires following the time evolution of
these four isotopes, as well as the knowledge of the associated $\beta$-spectra of their neutron-rich fission products.

This paper presents an improved treatment of the latter piece of information, common to the prediction of the $\nuebar$ 
spectrum of any moderated reactor. This work was triggered by the current effort of precision measurement of reactor neutrinos in 
the Double Chooz collaboration~\cite{Ardellier:2006mn}. Our approach combines the assets of the two main methods used so far. 
The first method is the so-called "{\it ab initio} approach" where the $\nuebar$ spectrum associated with one of the 4 fissioning 
isotopes is computed as the sum of the contributions from all fission products. This requires a huge amount of information on the 
thousands of $\beta$-branches involved and the weighting factors of fission products, the fission yields. Section~\ref{sec:ingredients} 
presents details on the ingredients of the {\it ab initio} approach while section~\ref{sec:micro_comp} gives an update of {\it ab initio} 
calculations combining all data available today. The main systematic errors are discussed and a prediction of $^{238}$U spectra 
is given since this isotope is the only one with no integral beta spectrum measured yet.

The second method relies on reference electron spectra~\cite{SchreckU5-1, SchreckU5-2, SchreckU5Pu9, SchreckPu9Pu1} measured 
at the high flux ILL reactor in Grenoble (France) using a high resolution magnetic spectrometer~\cite{BILL}. It is presented in 
section~\ref{sec:conversion} where we explain how these electron spectra are converted into antineutrino spectra with 
incomplete knowledge of the underlying physical distribution of $\beta$-branches. We show how our "mixed-approach" 
can improve the control of systematic errors and lead to a significant correction of the reference neutrino spectra used by 
all oscillation experiments so far. Finally we discuss in section~\ref{sec:ReactorNuExp} our results in the context of neutrino 
reactor experiments.

\section{Ingredients of Reactor Spectra}
\label{sec:ingredients}

In the present work we describe the total $\beta$ spectrum emitted by a reactor as the sum of the contributions from the
four fissioning nuclei mentioned in section~\ref{sec:intro}
\beqn\label{eq:Stot}
S_{\text{tot}}(E)&=&\sum_{k = ^{235}\text{U}, ^{238}\text{U}, ^{239}\text{Pu}, ^{241}\text{Pu}} \alpha_k\times S_{k}(E)
\eeqn
where $\alpha_k$ is the number of fissions of the $k^{\text{th}}$ isotope at the considered time,
$S_{k}(E)$ is the corresponding $\beta$ spectrum normalized to one fission and $E$ is the kinetic energy of emitted
electrons. 

Most of the equations below can be found in textbooks but they are useful here to define our notations and discuss the
systematic errors in the following sections. In the {\it ab initio} approach, $S_{k}(E)$ is broken up into the sum of
contributions from all fission products. 
\beqn\label{eq:Sk}
S_{k}(E)&=&\sum_{fp=1}^{N_{fp}}\mathcal{A}_{fp}(t)\times S_{fp}(E)
\eeqn
where $\mathcal{A}_{fp}(t)$ is the activity of the $fp^{\text{th}}$ fission product at time $t$ and normalized to one fission 
of isotope "$k$". Then the spectrum $S_{fp}(E)$ of each fission product is itself a sum of $N_b$ $\beta$-branches connecting 
the ground state (or in some cases an isomeric state) of the parent nucleus to different excited levels of the daughter nucleus
\beqn
\label{eq:Sfp}
S_{fp}(E)&=&\sum_{b=1}^{N_b} BR_{fp}^b\times S_{fp}^b(Z_{fp},A_{fp},E_{0fp}^b,E)
\eeqn
$BR_{fp}^b$ and $E_{0fp}^b$ are the branching ratio and the endpoint energy of the $b^{\text{th}}$ branch of the 
$fp^{\text{th}}$ fission product respectively. $Z_{fp}$ and $A_{fp}$ are the charge and atomic number of the parent 
nucleus. The sum of the branching ratios is normalized to the $\beta$-decay partial width of the parent nucleus 
($1$ if the parent is a pure $\beta^-$ emitter, $<1$ otherwise).

Equations (\ref{eq:Stot}) to (\ref{eq:Sfp}) are valid for both electron and antineutrino spectra. The expression of the
electron spectrum of the $b^{\text{th}}$ branch is given by the product of the following terms
\beqn\label{eq:Sfpb}
\nonumber S_{fp}^b & = & \underbrace{K_{fp}^b}_{\text{Norm.}} \times \underbrace{\mathcal{F}(Z_{fp},A_{fp},E)}_{\text{Fermi function}} \times \underbrace{pE(E-E_{0fp}^b)^2}_{\text{Phase space}} \\
& \times & \underbrace{C_{fp}^b(E)}_{\text{Shape factor}} \times \underbrace{\Big(1+\delta_{fp}^b(Z_{fp},A_{fp},E)\Big)}_{\text{Correction}}
\eeqn
To obtain the corresponding expression for the antineutrino spectrum one can safely neglect the nucleus recoil, and
replace in the  above formula the electron energy $E$ by the antineutrino energy 
\beqn\label{eq:Econserv}
E_\nu &=& E_{0fp}^b-E 
\eeqn
By definition this one-to-one relation is valid only at the single $\beta$-branch level. Thus this is a unique feature
of the  {\it ab initio} approach to predict electron and antineutrino spectra with the same precision.
The normalization factor $K_{fp}^b$ of Eq.(\ref{eq:Sfpb}) is calculated so that the integral $\int_0^{E_0}
S_{fp}^b(E)\,dE=1$. Hence the contribution to the integral of $S_{fp}(E)$ is driven by the branching ratio, as it should
be. The next two terms come from the Fermi theory. The Fermi function $\mathcal{F}(Z_{fp},A_{fp},E)$ corrects for the
deceleration of the electron in the Coulomb field created by the $Z_{fp}\times e$ positive charge of the parent nucleus.
Therefore in the case of $\beta^-$ decay the Fermi function causes the electron spectrum to start at a non zero value at
zero kinetic energy. This corresponds to a sharp step at the endpoint energy for the antineutrino spectrum, leading to
discontinuities when summing several branches of different endpoints.

The shape factor $C_{fp}^b(E)$ brings extra energy dependence beyond the trivial phase space factor of the Fermi theory,
due to the nuclear matrix element connecting the two nuclear levels of the $\beta$-decay. Its complexity depends on the
forbiddenness of the transition, driven by the spin-parity of the connected levels. In the case of allowed transitions
$C_{fp}^b(E)$ is a constant and is absorbed in the normalization factor.

For accurate predictions one must also take into account corrections, represented by the $\delta_{fp}^b$ factor in
Eq.(\ref{eq:delta}). This term is threefold
\beqn\label{eq:delta}
\nonumber \delta_{fp}^b(Z_{fp},A_{fp},E) = \delta_{QED}(E) & + & A_C(Z_{fp},A_{fp})\times E\\
 & + & A_W\times E
\eeqn
The $\delta_{QED}$ term corrects for radiation of real and virtual photons by the charged fermion lines of the
$\beta$-decay vertex. Its expression has been calculated at order $\alpha_{QED}$ by Sirlin {\it et al.}~\cite{Sirlin67}.
The fact that only the charged fermions radiate photons implies that the $\delta_{QED}$ formula differs for electron and
antineutrino spectra, the electron spectrum deviating more from the shape predicted by lowest order calculation than
that of the antineutrino. Strictly speaking, Eq.(\ref{eq:Econserv})  now becomes $E_0 = E_e + E_\nu + E_\gamma$ where
$E_\gamma$ represents the energy of the radiated photon. Still the $E_\gamma$ spectrum goes like $1/E_\gamma$ and the
dominant  contribution comes from soft ($E_\gamma\ll E_0$) radiated photons. Therefore the total energy of the lepton
pair remains very close to $E_0$. The physical constraint of conservation of the number of particles is fulfilled by the
equality
\beqn\label{eq:int_delta}
\nonumber \int_0^{E_0} S_{fp}^b&(&E_e) \times \big(1 +\delta_{QED}^e(E)\big)\,dE_e = \\
& & \int_0^{E_0} S_{fp}^b(E_\nu) \times \Big(1+\delta_{QED}^\nu(E)\Big)\,dE_\nu
\eeqn
which we verified numerically. The $A_C$ term is a Coulomb correction induced by the finite size of the decaying
nucleus. It is related to the interference of $<\vec\sigma>$ and $<\vec\sigma . \vec{r}^{\,2}/R^2>$ matrix elements, where R
is the nuclear radius and $\vec\sigma$ the spin operator. In the following we use the approximate expression derived by
Vogel~\cite{VogelPrivate}
\beqn\label{eq:A_C}
A_C &=& -\,\frac{10}{9}\,\frac{Z\alpha R}{\hbar c}
\eeqn
and the Elton formula~\cite{Gove} as an estimate of nuclear radii.

The nucleon itself also has a finite size and as a consequence the expression of its weak current deviates from that of
a point-like particle. The complex effects of the nucleon internal structure can be absorbed in the definition of factors
in front of each term of the most general nucleon weak current allowed by the symmetries of the theory. The equivalent in
the electromagnetic sector is the introduction of the Pauli ($F_1$) and Dirac ($F_2$) form factors of the nucleon, with
the $F_2$ contribution proportional to the momentum transfer. The $A_W$ term contains the CVC partner contribution of
$F_2$ in the vector weak current and is called the weak magnetism correction. Again we choose as a reference expression
the one derived by Vogel~\cite{VogelPrivate}
\beqn\label{eq:A_W}
A_W &=& -\frac{4}{3}\,\frac{\kappa_p-\kappa_n -1/2}{M_N\lambda}
\eeqn
with $\lambda=g_A/g_V=-1.2695$ the neutron disintegration constant and $M_N=939$ MeV the nucleon mass. One recognizes
the $(\kappa_p-\kappa_n)/M_N$ term proportional to $F_2$ at small ($\ll M_N^2$) square momentum transfer. Other
expressions of $A_C$ and $A_W$ can be found in the litterature~\cite{Holstein}. The associated uncertainty is large and
amplified by a sign compensation between the two terms. The net effect of these finite size corrections and its final
error are discussed in sections~\ref{sec:micro_comp} and~\ref{sec:conversion}.

\section{{\it ab initio} computation of Beta spectra from fission fragments}
\label{sec:micro_comp}

\subsection{Selection of the best data set}
\label{subsec:bestdata}

In principle the ultimate prediction of the $S_k(E)$ spectra comes from the knowledge of all quantities involved in
equations (\ref{eq:Sk}) to (\ref{eq:Sfpb}). The first attempts at such {\it ab initio} approach were
theoretical~\cite{GrossTheo,AvignoneIII, Vogel81,Klapdor82}. Most of these calculations use rather crude models to
describe the hundreds of involved nuclei but their goal is a correct description of total fission spectra $S_k(E)$, not
individual $\beta$-transitions. 

Efforts have also been put recently in comparing microscopic models, 
mostly theoretical models based on QRPA and the nuclear Shell Model, in the framework of double beta decay studies
\cite{Zuber}. Whereas these microscopic models are the ones susceptible to give the most reliable predictions, they are
still difficult to apply to large sets of nuclei, especially heavy nuclei (such as the large mass region of the fission
products) because of the large model spaces required. The estimation of the error associated to theoretical predictions
remains a difficult task and in practice they are supplanted by measurements performed  in the 80's at the ILL High Flux
Reactor in Grenoble~\cite{SchreckU5-1,SchreckU5-2,SchreckU5Pu9,SchreckPu9Pu1}. Only the $^{238}$U spectrum remained 
calculated~\cite{CalcU8,Vogel81} since no related data exist yet. Nevertheless a measurement in the fast neutron flux of the 
FRMII reactor in Garching has lately been performed~\cite{Nils} and should be published soon.

We describe here a complementary {\it ab initio} approach with the strategy of exploiting all data available in modern
nuclear databases while reducing the input of nuclear models. The total spectrum $S_k(E)$ of each fissioning isotope is
built up according to the equations of section~\ref{sec:ingredients}, retrieving the information on all $\beta$-branches
from the ENSDF nuclear database~\cite{ENSDF}. The motivations for such an approach are that when all parameters of a
$\beta$-branch are known the neutrino  branch is also known in a model-independent way and all errors on the input
parameters can in principle be propagated. We have developed an interface with the ENSDF data library to read the
relevant parameters of Eqs. (\ref{eq:Sfp}) and (\ref{eq:Sfpb}) and their experimental error. The forbiddenness of a
$\beta$-transition is deduced from the spin and parity of the connected nuclear levels. In cases when this information
is missing or uncomplete, the lowest possible forbiddenness is chosen by default. All transitions tagged as forbidden
are then forced to be of unique type and the corresponding expressions of the shape factors $C_{fp}^b(E)$ are
polynomials in the electron and antineutrino momenta taken from~\cite{Behrens}. Using our homemade code BESTIOLE, 
all the above approximations used for the calculation of each branch are tagged and various scenarios can be tested to 
estimate the error envelope of the final predicted spectrum. From Eqs.(\ref{eq:Sfp}) and (\ref{eq:Sfpb}) the electron and 
antineutrino spectra of each fission product are computed and stored in a database.  

Then the total beta spectrum of one fissioning isotope is built as the sum of all fission
fragment spectra weighted by their activity (Eq.\ref{eq:Sk}). These activities are determined using a simulation
package called MCNP Utility for Reactor Evolution (MURE~\cite{MURE}). MURE is a precision code written in C++ which
automates the preparation and computation of successive MCNP (Monte-Carlo N-Particle transport code~\cite{MCNP})
calculations either for precision burn-up or thermal-hydraulics purpose. It is open-source, portable, and available at
NEA \cite{MURENEA} and constitutes an efficient tool for non-proliferation and thermal power scenario studies (for more details see
\cite{ProceedingsNonProlif}).

\begin{figure}[t]
\includegraphics[width=0.9\linewidth]{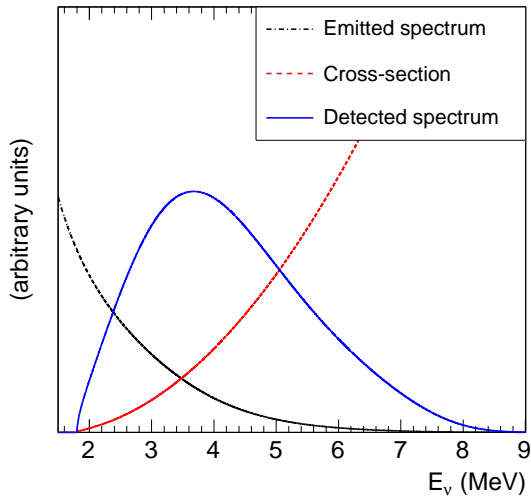}
\caption{(Color online) Illustration of the detected antineutrino spectrum in the case of $^{235}$U fissions (blue solid curve). Units are arbitrary 
and oscillation effects are suppressed. The detected rate rises from the threshold value at about 1.8 MeV, reaches a maximum 
around 4 MeV and vanishes after 8 MeV. This shape is the result of folding the emitted spectrum (black dashed-dotted curve),  parameterization 
taken from section \ref{subsec:Param} and beta-inverse cross section (red dashed curve).}
\label{fig:DetSpec}
\end{figure}

The detection process, common to many reactor antineutrino experiments, is the $\beta$-inverse reaction on a free proton
\beqn\label{eq:beta_inv}
\nuebar + p \ra e^+ + n
\eeqn
which sets an energy threshold for the antineutrino of 1.804 MeV, the mass difference between the initial and final
states (see figure \ref{fig:DetSpec}). Therefore the lowest energy part of the $\beta$ spectra, below this threshold, is
not addressed here. In this region, equilibrium is reached only after several months, requiring the control of
significant transient effects when considering shorter irradiation times. Extra effects like the low energy
$\beta$-decays induced by neutron capture on $^{238}$U and fission products~\cite{Kopeikin} would also have to be
treated. On the high energy side, antineutrino rates above 8 MeV become negligible ($< 0.5\%$ of total detected rate).
This part of the spectrum is dominated by the very energetic (high $Q_\beta$) transitions of rare exotic nuclei and
cannot be accurately predicted. Thus the intermediate energy range resulting from the observation of the detected
spectrum in figure \ref{fig:DetSpec} turns out to be favorable to the control of the systematic errors of the
predictions of reactor antineutrino spectra.

\begin{figure}[t]
\includegraphics[width=0.9\linewidth]{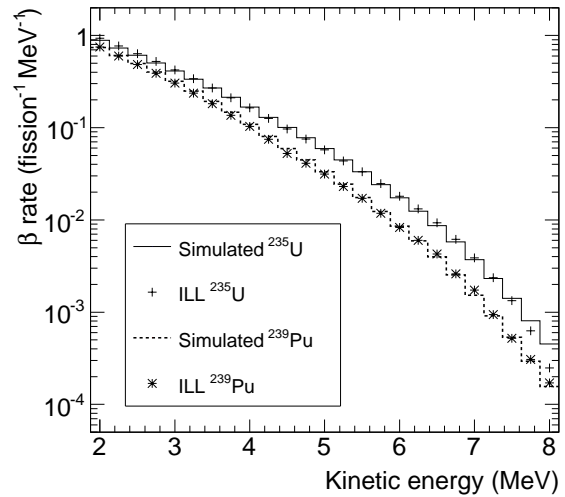}
\caption{Comparison of $^{235}$U and $^{239}$Pu reference electron spectra from~\cite{SchreckU5Pu9} with our predictions based 
on the {\it ab initio} approach. The predictions have no free parameters and the rates are normalized to one fission.}
\label{fig:U235_abs}
\end{figure}

A powerful test of our calculations is the comparison with the reference electron spectra from 
ILL~\cite{SchreckU5-2,SchreckU5Pu9,SchreckPu9Pu1}.  Such a consistency
check gives valuable insight into the distribution of the numerous $\beta$-branches, pointing to the main source of
errors in the determination of the antineutrino spectra. Considering all the data available in the ENSDF data library,
the predicted $\beta$-spectra associated with the fission of $^{235}$U and $^{239}$Pu are compared with the reference
ILL data in figure~\ref{fig:U235_abs}.

\begin{figure}[h]
\includegraphics[width=0.85\linewidth]{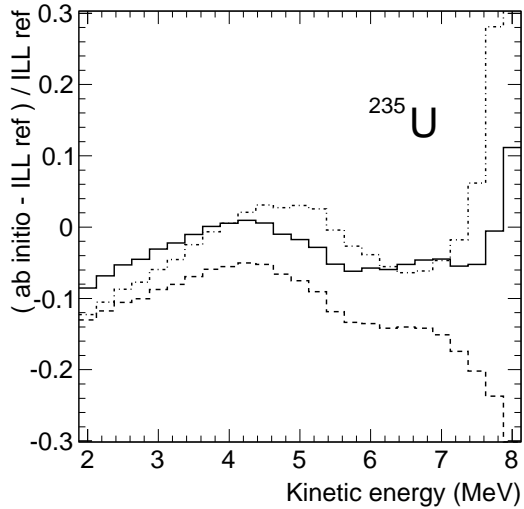}
\caption{Residues of the $^{235}$U electron spectra computed as the difference of our {\it ab initio} calculations minus 
reference data from~\cite{SchreckU5Pu9} divided by reference data. Dashed-dotted curve: ENSDF data only; dashed curve: 
some ENSDF data replaced by pandemonium corrected data; solid curve: unmeasured $\beta$ emitters are added on top 
of previous curve, using the gross-theory calculations of the JENDL nuclear database and few remaining exotic nuclei described 
by our model (see text).}
\label{fig:U235_AbInit}
\end{figure}

Although the spectrum falls quickly with energy, reasonable agreement is found on the shape and absolute normalization
over a quite large energy range. Note that our prediction is parameter free. For finer analysis the residues of our
predicted $^{235}$U spectrum with respect to reference data are shown as the dashed-dotted line in
figure~\ref{fig:U235_AbInit}. It reveals a $\pm$ 10\% oscillation pattern of the calculations around the data up to 7.5
MeV and a large overestimation at higher energy. This overestimation points to the well known systematic effect of
pandemonium~\cite{Pande}. Indeed branching ratios and endpoints are usually determined by measuring the intensity and
energy of $\gamma$-radiations emitted subsequently to the $\beta$ transition using high resolution but low efficiency Ge
crystals. In the case of large $Q_\beta$ a fraction of the beta branches connects the parent nucleus to very excited
levels of the daughter nucleus. The strength of the associated low energy $\beta$-rays is either spread over multiple
weak $\gamma$-rays or concentrated in one high energy gamma ray. In both cases part or all the $\gamma$-cascade can be
missed by the measurement apparatus. As a result low endpoint transitions are often missed and high endpoints are given
too much weight in the global decay scheme of the parent nucleus.

To correct for the pandemonium effect we tried to gather $\beta$-decay data using other experimental techniques than the
$\beta - \gamma$ coincidence. A valuable set of data comes from the measurement campaign undertaken by Tengblad {\it et
al.}~\cite{Tengb} in the late eighties at the on-line isotopes separators ISOLDE, at CERN, Geneva, and OSIRIS at the
Neutron Research Laboratory, Studsvik. Some 111 fission products, selected as the main contributors to the high energy
part of reactor beta spectra (90\% above 6 MeV) were measured. Electron spectra were recorded independently from the
emitted gamma rays. This prevented sensitivity to the pandemonium effect but at the same time part of the information on
single $\beta$-branches was lost. Among the 111 measured electron spectra, 44 were found in perfect agreement with the
spectra predicted from the ENSDF database. The remaining 67 were then replaced in our database.

Another important source of data are the measurements based on Total Absorption Gamma Spectrometers (TAGS). The
principle here is to implant the radioactive isotope on a foil surrounded by high efficiency gamma detectors able to
collect the whole $\gamma$-cascade following the beta-decay. The distribution of total $\gamma$ energies gives access to
the beta-strength of the studied isotope at the cost of a deconvolution analysis taking into account the full response
of the apparatus. Eventually a complete beta-decay scheme can be determined providing the relevant beta-branch
information for electron and antineutrino spectra. Thus the 29 nuclei of R. Greenwood {\it et al.}~\cite{Green} measured
at the INEL facility, Idaho, were incorporated in our database. A.~Algora and J.~L.~Tain studied carefully both Tengblad
et al. and Greenwood et al.'s measurements~\cite{PrivateComTain}. Both may be affected by several sources of systematic
effects which are difficult to quantify. In particular both measurements of $^{91}$Rb beta decay mean energy differ by
more than 350 keV, while the $^{91}$Rb decay scheme was used in Tengblad et al.'s analysis to quantify the $\gamma$-ray
detector absolute efficiency. If the $^{91}$Rb decay scheme is affected by the pandemonium effect, Tengblad et al.'s
data sets may exhibit an overall systematic effect. A new TAGS measurement of the decay properties of $^{91}$Rb has
recently been performed at the Jyv\"{a}skyl\"{a} University facility \cite{tain2010}, and will help addressing the
uncertainties associated to both sets of measurements. In cases when a fission product was present in both data sets (8
nuclei only), giving the priority to Greenwood {\it et al.}'s or Tengblad {\it et al}'s measurements changes the
predicted spectrum by 3\% at most in the 4-5 MeV range, the effect drops at the 1\% level or below elsewhere. In the
following priority is arbitrarily given to Tengblad {\it et al}'s data. The dashed line in figure~\ref{fig:U235_AbInit} shows 
the electron residues after merging our ENSDF based database with the above selected spectra supposedly corrected for 
the pandemonium effect. As expected the high energy part of our prediction has been
significantly reduced leading to negative residues of increasing amplitude with energy. This indicates that a large part
of the pandemonium effect is probably corrected and that contributions from the missing unknown transitions of exotic
nuclei grow rapidly with energy. 

To fill up this missing contribution we collected all available predictions of electron spectra from the JENDL nuclear
database \cite{JENDL}. These predictions are based on the "Gross Theory of Beta Decay" \cite{GrossTheo2} and were 
included in the JENDL database to supplement ENSDF data showing incomplete level schemes or
for nuclei for which no data were available. The estimated spectra were stored in the JENDL file so as to keep the
consistency between the average decay energy value derived from the spectrum and that used for decay heat analysis
\cite{GrossTheo3}. The calculated spectra were also compared with the directly measured spectra from the reference
\cite{Tengb} and revealed to be in very good agreement. The total contribution of the JENDL electron spectra, not
already included in the ENSDF and pandemonium corrected data, was computed and converted to its associated total
antineutrino spectrum following the procedure described in \cite{SchreckU5Pu9}. Finally the few remaining nuclei were
described using a model based on fits of the distributions of the end-points and branching ratios in the ENSDF database,
then extrapolated to the exotic nuclei. The result is the solid curve in figure~\ref{fig:U235_AbInit} showing
flattened residues within a global envelope of $\pm$ 10\% over the whole energy range. This best agreement with the ILL
reference data is actually valid for the $^{239}$Pu isotope. Depending on the considered fissionning isotope, this new
compilation of $\beta$-decay data includes about 845 nuclei and 10000 $\beta$-branches, about 525 nuclei come from the
ENSDF and pandemonium corrected data, 285 from the JENDL database and 35 from our model. It represents the best data set
for {\it ab initio} calculation.

\subsection{Results}
\label{subsec:results}

From this work we conclude that a compilation of all available data on the beta-decays of fission products can
describe the antineutrino fission spectra at the 10\% level, illustrating the tremendous experimental work already
achieved. Still, the relatively large energy range of detected antineutrino involves a sizeable contribution of unstable
and poorly known nuclei in the total spectrum. Under these conditions, improving errors or even reaching the accuracy of
the ILL reference spectra seems to require another fair amount of experimental effort. For applications like the
determination of reactors decay heat calculations, a short  list of "pandemonium candidates" to be remeasured with total
absorption techniques has been identified~\cite{Yosh99}. Completion of a corrected beta-decay database is in progress (see
for instance~\cite{Algora}) with more and more refined analyses~\cite{Tain}. Thanks to our database of fission product
spectra, we have established a list of nuclei, contributing importantly to different energy bins of the antineutrino
energy spectra from $^{235}$U and $^{239}$Pu, and that could be affected by the pandemonium effect. It will be the 
subject of a forthcoming publication. From this list we have selected a few nuclei which are amenable to experimental 
investigation using the TAGS technique, which can provide the beta
intensity distribution in the full decay window eliminating the pandemonium effect. It appeared that some fission
products being part of the measurement priority list selected for reactor decay heat assessment \cite{Algora}, also
belong to the list of important contributors to the antineutrino emission in the energy window of interest for neutrino
oscillation studies. Recent and on-going experimental efforts carried out in the field of reactor physics, neutrino
physics but also of interest for nuclear structure and astrophysics will certainly allow to reduce the uncertainties
associated to the reactor antineutrino spectra computed through the ab-initio method in the very next years
\cite{proposals, tain2010, algora2010}. We describe below the estimated error budget of our {\it ab initio} calculations
and give a prediction of electron and antineutrino spectra of $^{238}$U. In the perspective of neutrino oscillation
analyses the  $^{235}$U and $^{239}$Pu isotopes, which contribute to about 90\% of a nuclear reactor spectrum, are
predicted using a more accurate method presented in section \ref{sec:conversion}.

\subsection{Error Budget}
\label{subsec:errorbudget}

As mentioned earlier, the control of the parameters of all single $\beta$-branches allows a full propagation of the
errors quoted in the ENSDF database. All sources of error are treated as independent and the total error matrix of rates
in energy bins is computed. In the simpler case of a spectrum at equilibrium, the activity of each fission product is
approximated by the associated cumulative fission yield indexed in the JEFF3.1.1 database~\cite{JEFF}. Then the
uncertainty on branching ratios and fission yields can be propagated analytically while the uncertainty on end-points is
propagated numerically (it turns out to have a negligible contribution). The dominant contribution of normalization
errors induces large correlations between proximate bins as illustrated in table~\ref{tab:CorrelMatrix}. Note that these
correlations are valid only for the specific part of the measurement errors quoted in the nuclear databases. As
summarized in table~\ref{tab:micro_errors} we know from the above section that systematic effects beyond these databases
are dominant and will change these correlations in a non-trivial way as long as all $\beta$-branches are not corrected.

\begin{table}[t]
\begin{center}
\medskip
\begin{tabular}{c}
 $\rho = \left(
  \begin{array}{ccccccc}
       1 & 0.54 & 0.48 & 0.41 & 0.38 & 0.34 & \dots \\
   0.54 &     1 & 0.48 & 0.43 & 0.39 & 0.35 & \\
   0.48 & 0.48 &     1 & 0.46 & 0.42 & 0.38 & \\
   0.41 & 0.43 & 0.46 &     1 & 0.42 & 0.39 & \\
   0.38 & 0.39 & 0.42 & 0.42 &     1 & 0.39 & \\
   0.34 & 0.35 & 0.38 & 0.39 & 0.39 &     1 & \\
  \dots & & & & & & \\
   \end{array}\right)$
\end{tabular}
\caption{Correlation matrix in the range 2 to 3.5 MeV in 250 keV bins obtained by propagating all sources of errors in ENSDF 
and JEFF databases. Branching-ratio errors cause a very high level of correlation then reduced by the end-point distribution and 
assumption of independent fission yields. \label{tab:CorrelMatrix}}
\end{center}
\end{table}

\begin{table}[h]
\begin{center}
\medskip
\begin{tabular}{|c|c|c|c|c|}
\hline\hline
Kinetic & Nuclear   & Forbid. & A$_{\text{C,W}}$ & Missing     \\
$E$ (MeV)       & databases & treatment     & corrections      & info. \\
\hline
2.00        &  1.2     &  0.2          &  0.1             &  10      \\
2.25        &  1.3     &  0.2          &  0.2             &  10      \\
2.50        &  1.3     &  0.1          &  0.3             &  10      \\
2.75        &  1.3     &  0.1          &  0.3             &  10      \\
3.00        &  1.4     &  0.4          &  0.4             &  10      \\
3.25        &  1.6     &  0.7          &  0.5             &  10      \\
3.50        &  1.7     &  0.1          &  0.5             &  10      \\
3.75        &  1.9     &  1.3          &  0.6             &  10      \\
4.00        &  2.2     &  1.6          &  0.6             &  10      \\
4.25        &  2.5     &  1.6          &  0.7             &  10      \\
4.50        &  2.8     &  1.4          &  0.8             &  10      \\
4.75        &  3.2     &  1.0          &  0.8             &  10      \\
5.00        &  3.8     &  0.5          &  0.9             &  10      \\
5.25        &  4.4     &  0.2          &  0.9             &  10      \\
5.50        &  5.2     &  0.2          &  0.9             &  15      \\
5.75        &  6.1     &  0.2          &  0.9             &  15      \\
6.00        &  7.1     &  0.2          &  1.0             &  15      \\
6.25        &  8.0     &  0.3          &  1.0             &  15      \\
6.50        &  9.0     &  0.4          &  1.1             &  15      \\
6.75        &  10.1     &  0.4          &  1.1             & 15      \\
7.00        &  10.9     &  0.5          &  1.1             & 20      \\
7.25        &  11.0     &  0.7          &  1.1             & 20      \\
7.50        &  10.7     &  0.8          &  1.1             & $>20$      \\
7.75        &  11.1     &  0.8          &  1.2             & $>20$      \\
8.00        &  13.3     &  1.2          &  1.3             & $>20$      \\
\hline\hline
\end{tabular}
\caption{Sources of errors in the $^{235}$U electron spectrum as predicted by the {\it ab initio} approach. All errors are given 
in percent at 1$\sigma$ (68\% CL). \label{tab:micro_errors}}
\end{center}
\end{table}

The second column of table~\ref{tab:micro_errors} lists the global effect of the errors quoted in ENSDF at the 1 sigma
level. It rises from 1 to 10 \% in the 2-8 MeV range. Columns 3 and 4 show the impact of the theoretical assumptions
used to describe the shape of the $\beta$-branches. Previous works always treated all branches as allowed. Comparing
this hypothesis with our full treatment of forbiddenness shows changes of the final antineutrino spectrum below the
$1\%$ level (except for few bins around 4 MeV), validating the allowed approximation. The error associated
with the finite size corrections A$_C$ and A$_W$ has been estimated by comparing the final spectra computed with no
correction and those with the corrections from Vogel~\cite{VogelPrivate} or Holstein~\cite{Holstein}. The missing
information on exotic nuclei and the correction of the pandemonium effect unfortunately remain the dominant contribution
in the final error of the {\it ab initio} approach. It is roughly estimated in the last column of the table based on the
envelope of the various scenarios we tried and on the residues with respect to the reference ILL data.

\begin{figure}[h]
\includegraphics[width=0.9\linewidth]{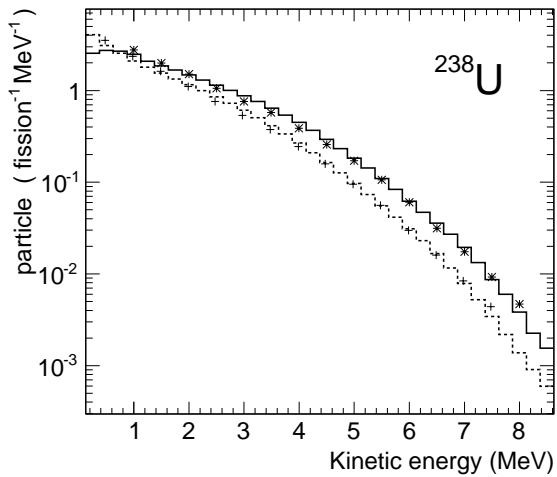}
\caption{{\it ab initio} calculation of the electron (dashed histogram) and antineutrino (solid histogram) spectra of $^{238}$U 
for a 450 days irradiation time. For comparison the predictions of \cite{Vogel81}  for an infinite irradiation time are plotted as crosses 
for the electron spectrum and stars for the antineutrino spectrum. \label{fig:238U_AbInit}}
\end{figure}

\subsection{Predictions of $^{238}$U spectra}
\label{subsec:238Uspectra}

\begin{table}[t]
\begin{center}
\medskip
\begin{tabular}{|c|c|c|c|c|}
\hline\hline
Kinetic $E$ & \multicolumn{2}{c|}{$N_\beta$}  & \multicolumn{2}{c|}{$N_{\bar\nu_e}$} \\
(MeV)       & \multicolumn{2}{c|}{(/fission/MeV)} & \multicolumn{2}{c|}{(/fission/MeV)} \\ 
 & 12 h & 450 d & 12 h & 450 d \\
\hline
2.00  &  1.13          &  1.15          &  1.43       &  1.48       \\
2.25  &  9.81$(-1)$&  9.97$(-1)$&  1.26       &  1.30       \\
2.50  &  8.47$(-1)$&  8.55$(-1)$&  1.12       &  1.15       \\
2.75  &  7.24$(-1)$&  7.27$(-1)$&  9.80$(-1)$&  1.00     \\
3.00  &  6.11$(-1)$& 6.11$(-1)$&  8.70$(-1)$&  8.76$(-1)$ \\
3.25  &  5.07$(-1)$&  5.06$(-1)$&  7.57$(-1)$&  7.59$(-1)$ \\
3.50  &  4.16$(-1)$&  4.15$(-1)$&  6.40$(-1)$&  6.42$(-1)$ \\
3.75  &  3.37$(-1)$&  3.36$(-1)$&  5.39$(-1)$&  5.39$(-1)$ \\
4.00  &  2.68$(-1)$&  2.67$(-1)$&  4.50$(-1)$&  4.51$(-1)$ \\
4.25  &  2.11$(-1)$&  2.10$(-1)$&  3.67$(-1)$&  3.67$(-1)$ \\
4.50  &  1.64$(-1)$&  1.63$(-1)$&  2.94$(-1)$&  2.93$(-1)$ \\
4.75  &  1.27$(-1)$&  1.27$(-1)$&  2.32$(-1)$&  2.32$(-1)$ \\
5.00  &  9.72$(-2)$&  9.69$(-2)$&  1.83$(-1)$&  1.83$(-1)$ \\
5.25  &  7.37$(-2)$&  7.33$(-2)$&  1.43$(-1)$&  1.43$(-1)$ \\
5.50  &  5.55$(-2)$&  5.52$(-2)$&  1.10$(-1)$&  1.10$(-1)$ \\
5.75  &  4.17$(-2)$&  4.14$(-2)$&  8.35$(-2)$&  8.35$(-2)$ \\
6.00  &  3.12$(-2)$&  3.10$(-2)$&  6.21$(-2)$&  6.21$(-2)$ \\
6.25  &  2.31$(-2)$&  2.30$(-2)$&  4.70$(-2)$&  4.70$(-2)$ \\
6.50  &  1.68$(-2)$&  1.66$(-2)$&  3.58$(-2)$&  3.58$(-2)$ \\
6.75  &  1.17$(-2)$&  1.16$(-2)$&  2.71$(-2)$&  2.71$(-2)$ \\
7.00  &  7.92$(-3)$&  7.85$(-2)$&  1.95$(-2)$&  1.95$(-2)$ \\
7.25  &  5.28$(-3)$&  5.23$(-2)$&  1.32$(-2)$&  1.33$(-2)$ \\
7.50  &  3.48$(-3)$&  3.44$(-2)$&  8.65$(-3)$&  8.65$(-3)$ \\
7.75  &  2.22$(-3)$&  2.19$(-2)$&  6.01$(-3)$&  6.01$(-3)$ \\
8.00  &  1.40$(-3)$&  1.38$(-2)$&  3.84$(-3)$&  3.84$(-3)$ \\
\hline\hline
\end{tabular}
\caption{$^{238}$U electron and antineutrino spectra obtained by combining our best compilation of data sets with the activity 
of all fission products as predicted by the MURE evolution code after a 12 h and 450 days irradiation time. Associated errors are those listed in 
table~\ref{tab:micro_errors}. \label{tab:238U_spectra}}
\end{center}
\end{table} 

As no experimental data on $^{238}$U are available at the present time we provide a prediction of its electron and
antineutrino spectra using the above {\it ab initio} approach. This calculation has been performed using our best data
set as defined in section \ref{subsec:bestdata}. The $^{238}$U spectrum is given in table~\ref{tab:238U_spectra} after 
two irradiation periods into a neutron flux: 12 h, similar to the irradiation time of the ILL data and 450 days as an approximation 
of a spectrum at equilibrium. All neutron captures effects are turned off for this prediction and the error budget described in 
table~\ref{tab:micro_errors} applies to this prediction. Comparison with a previous estimate~\cite{Vogel81} is illustrated in 
figure \ref{fig:238U_AbInit}. Despite some slight differences in shape, our work and previous predictions agree within $\pm 10\%$ 
across the full energy range. After multiplication with the $\beta$-inverse cross section (Eq.\ref{eq:beta_inv}), the net effect 
on the integrated detected neutrino flux is a 9.8\% increase for the $^{238}$U contribution.

\section{Improved Conversion of Reactor Electron Spectra into Antineutrino}
\label{sec:conversion}

In the previous section we showed that the {\it ab initio} approach had strong limitations due to unknown contribution
from very unstable nuclei. Nevertheless we keep in mind that the beta-transitions described in nuclear databases
represent about 90\% of the total spectrum as measured at ILL. These physical distributions of endpoints and nuclear
charges are precious information to control the conversion between electron and antineutrino spectra. We describe below
how this can be combined to the very precise ILL electron data for an improved prediction of antineutrino spectra.

\subsection{Previous conversion procedure}
\label{subsec:PrevConv}

The measurements performed at ILL gave access only to the global electron spectrum of a fissile isotope, {\it i.e.} the 
sum of the contributions of all fission products. Thin target foils of fissile isotopes $^{235}$U,
$^{239}$Pu and $^{241}$Pu  were exposed to the thermal neutron flux 80 centimeters away from the center of the compact
fuel assembly. A tiny part of the emitted electrons could exit the reactor core through a straight vacuum pipe to be
detected by the high resolution magnetic spectrometer BILL~\cite{BILL}. The electron rates were recorded by a point wise
measurement of the spectrum in magnetic field steps of 50 keV, providing an excellent determination of the shape of the
electron spectrum with sub-percent systematic error. The published data are smoothed over 250 keV. Except for the
highest energy bins with poor statistics, the dominant error was the absolute normalization, quoted around 3\% (90\% CL)
with weak energy dependence. Note also that the ILL spectra are taken after typically 1 day of irradiation, meaning that the 
longest lived beta emitters (lowest energy beta rays) haven't reached equilibrium yet. These aspects are discussed in detail in 
section~\ref{subsec:off-eq}. \\
The neutrino spectra, not directly detected, were deduced from those of the electron via a conversion procedure which induced 
some extra systematic effects. In~\cite{SchreckU5-1,SchreckU5-2, SchreckU5Pu9, SchreckPu9Pu1} the authors considered 
30 virtual beta branches.
The procedure consisted in dividing the electron spectrum into 30 slices. Starting with the highest energy slice, the
few data points in this slice were used to fit the endpoint and branching ratio of the first virtual branch. The full
contribution of this virtual branch (from endpoint down to zero energy) was then subtracted from the experimental
spectrum and the procedure repeated for the next, lower energy, slice. Then the antineutrino spectrum was simply the sum
of all fitted virtual branches, converted to antineutrino branches by replacing $E_e$ by $E_\nu =E_0-E_e$ and applying
the correct radiative corrections. This procedure was repeated several times, describing the spectrum with somewhat
different sets of end point energies. Possible steps in the shape induced by the relatively small number of virtual branches
were smoothed out by taking the average of all spectra and merging the 50 keV bins into the 250 keV presented in the
publications. The theoretical expression of a virtual branch was the same as Eq.(\ref{eq:Sfpb}), except for the $A_C$
and $A_W$ corrections which were treated at the very end as an effective linear correction to the final antineutrino
spectra
\beqn\label{eq:eff_delta}
\Delta N_\nu^{WC} (E_\nu) &\simeq& 0.65\left(E_\nu-4.00\right)\,\%
\eeqn
with $E_\nu$ in MeV. The final error of the conversion procedure was estimated to be 3-4\% (90\% CL), to be added in
quadrature with the electron calibration error.

In a recent paper~\cite{Vogel08}, P. Vogel pointed out the main limitations of this conversion procedure. Despite the
discontinuity of the Fermi function at the endpoint energy of an antineutrino branch, the "true" antineutrino spectrum
from fission fragment appears continuous because thousands of branches contributes with a quasi-continuous endpoint
distribution. When describing the spectrum by only 30 virtual branches a spurious oscillation with respect to the true
spectrum appears around each virtual endpoint energy. Therefore smoothing out these oscillations requires sufficiently
narrow slices of electron data and antineutrino energy bins several times larger than the slice width. All these
criteria couldn't be fulfilled for the electron data taken at ILL and the estimation of the remaining effects is part of
the quoted error bar. The other criterion highlighted by P. Vogel was the knowledge of the average nuclear charge $<Z>$
of the virtual branches as a function of their endpoint energy. This information turns out to be of crucial importance
for the shape of the high energy part of the antineutrino spectrum.

\subsection{Improvements of the conversion procedure}
\label{subsec:ImprovedConv}

Our new conversion method allows us to address the sources of errors in a complementary way. It consists in starting
with our {\it ab initio} prediction of section~\ref{sec:micro_comp} and restrict the use of effective branches to fit
only the missing few percent contribution of the difference with the reference ILL electron data. This way we keep the
distributions of beta branches very close to the physical one and we can apply $A_C$ and $A_W$ corrections at the branch
level. The reference ILL electron data are still fitted but the contribution of unphysical virtual branches is reduced
by an order of magnitude. We use all available data in ENSDF plus the above-mentioned 67 nuclei from the "pandemonium
corrected" measurements. In the case of the data from reference~\cite{Tengb} only the total $\beta$ spectra of each
nuclide are available, not the complete decay scheme as would be best. Hence, to be converted to a neutrino spectrum,
each $\beta$ spectrum measured must be fitted by a set of branches. These branches differ from the virtual branches used
to fit the ILL data by the fact that in this work the nuclear charge is perfectly known. Moreover the validation of the
spectrum shape at the level of one nucleus provides a more refined description of the $\beta$-decay scheme equivalent to
a small slice size in the total spectrum, pinning down the sources of errors discussed in~\cite{Vogel08}.

\begin{figure}[!h]
\begin{center}
\includegraphics[width=0.9\linewidth]{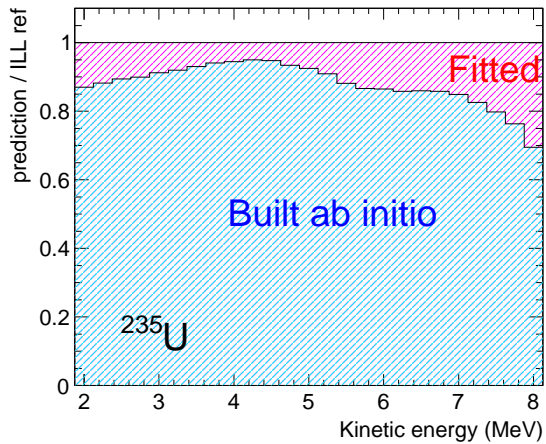}
\caption{(Color online) The blue hatched area shows the contribution of our {\it ab inito} prediction (ENSDF + pandemonium corrected nuclei)  
relative to the ILL reference data. The missing contribution coming from unknown nuclei and remaining systematic effects of 
nuclear databases (red hatched area) is fitted using a set of 5 effective $\beta$-branches. }
\label{fig:virtual_fit}
\end{center}
\end{figure}

\begin{figure}[!h]
\flushleft\includegraphics[width=.85\linewidth]{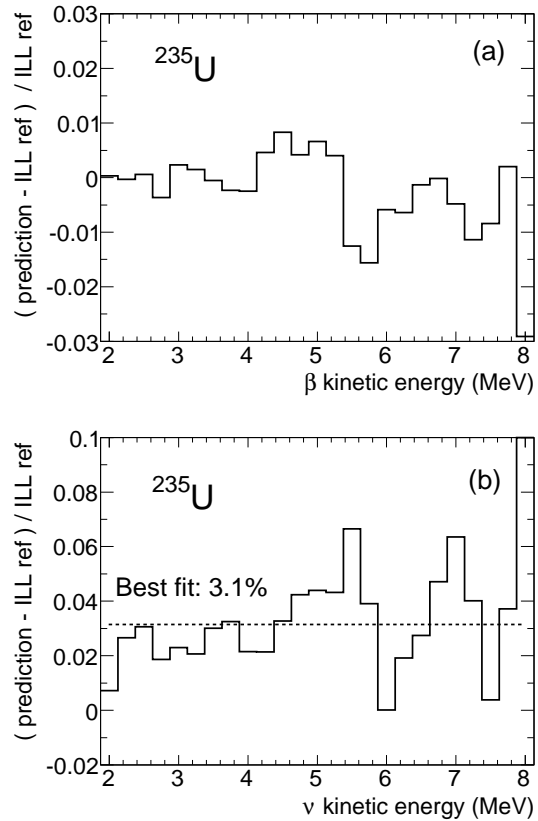}
\caption{Relative difference to ILL spectra for electron (a) and antineutrino (b) spectra for $^{235}$U. While electron 
residues are zeroed at the $\pm 1\%$ level by the fitting procedure, antineutrino residues exhibit a mean 
normalization shift of  about 3\% (dashed line in(b)).}
\label{fig:new_residues}
\end{figure}

On top of the contributions of all ENSDF and pandemonium corrected $\beta$-branches, the missing contribution to match
the ILL electron spectrum is fitted using a set of 5 effective $\beta$-branches with a nuclear charge of $Z$=46 (chosen
as the average of the distribution of fission products), and assuming that transitions are allowed transitions. The
normalization and the end-point are two free parameters for each branch. An example of the "{\it ab initio}" and
"fitted" contributions of $^{235}$U electron spectrum is illustrated in figure~\ref{fig:virtual_fit} as stacked
histograms. The residues of the fitted missing contribution are shown in figure~\ref{fig:new_residues}(a).
They are small, typically at the level of the statistical error of the ILL data, except in the 4.5-6.0 MeV range where
one can see an oscillation pattern with an amplitude reaching three times the error of the ILL data at maximum. This may
point to a systematic effect due to a failure of the fit model. We checked that using more effective branches is not
efficient because of the limited number of experimental points available. The impact of these non statistical residues
in the final error is discussed latter. 

\subsection{New reference antineutrino spectra}
\label{subsec:NewRef}

Converting all branches from the nuclear databases plus the 5 fitted ones into antineutrino branches (as described in
section~\ref{sec:ingredients}), we obtain the predicted antineutrino spectrum. The residues with respect to the
prediction of Schreckenbach {\it et al.} are shown in figure~\ref{fig:new_residues}(b). It exhibits a
good agreement in shape but a mean normalization shift of about +3\%. This shift of the emitted antineutrino flux is
modulated at higher energy by oscillations which look like images of the oscillations in the electron residues. When
folded with the $\beta$-inverse cross section the predicted increase of detected antineutrino rate from $^{235}$U is 
about 2.5\%. Note that this positive shift is a mean value computed between the energy threshold (1.8 MeV) imposed 
by the detection process and infinity. The physical constraint of one emitted electron for one emitted antineutrino must 
still be fulfilled. We did check that the total integral of our electron spectrum fitted on the ILL reference and the integral
of the associated converted neutrino spectrum were identical at the $10^{-4}$ level. Since all our individual $\beta$
and antineutrino branches are normalized to the same integral with much higher accuracy this result gives an estimate of
the numerical precision in the sum of thousands of branches.

\begin{figure}[t]
\includegraphics[width=0.9\linewidth]{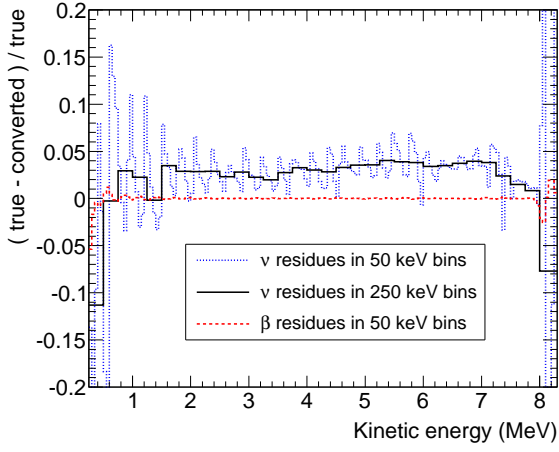}
\caption{(Color online) Independent cross-check of our results based on known reference spectra (pure ENSDF database). Dashed 
red line: electron residues after fitting with 30 virtual branches. Dotted blue line: relative difference between the reference 
antineutrino spectrum and the one converted according to the ILL procedure,  in 50 keV bins. Smoothing out the residual 
oscillations in 250 keV bins (solid black line) exhibits the same $3\%$ normalization shift than in figure~\ref{fig:new_residues}(b).}
\label{fig:TestConv}
\end{figure}

To  test the validity of our procedure we applied it to effective calculated electron and antineutrino spectra. 
This method was inspired from the work of P. Vogel~\cite{Vogel08}. We generated
electron and antineutrino spectra as the sum of the spectra of all fission products indexed in ENSDF, weighted by the
activity predicted by the MURE code after 12 hours of irradiation. We know from section~\ref{sec:micro_comp} that these
spectra are close to the ones measured at ILL and in the context of this test we call them "true" spectra in the sense
that they are unambiguously connected to each other by the conversion of each single branch of the sum. Then we followed
the exact same procedure as the one described in~\cite{SchreckU5-2,SchreckU5Pu9,SchreckPu9Pu1} to convert our "true"
electron spectrum into an antineutrino spectrum. This includes using 30 virtual branches and the same effective Z
distribution and the same effective $A_C+A_W$ correction (Eq.(\ref{eq:eff_delta})). The spectrum converted in this
fashion is finally compared to our "true" antineutrino spectrum. Figure~\ref{fig:TestConv} shows that despite a very
good fit quality of the electron spectrum (all electron residues are within a few $10^{-3}$ from 1 to 8 MeV) the
converted antineutrino spectrum exhibits residues with oscillations of few percent amplitude around the endpoint of each
fitted branch. As expected, rebinning smoothes out these oscillations (solid curve) but a residual $\simeq +3\%$ offset
is clearly visible across the whole energy range. This curve can be directly compared to the result of our conversion of
the ILL data, in figure~\ref{fig:new_residues}(b). Very good agreement is found, validating the above predicted deviation 
from the ILL antineutrino spectra. 

\begin{figure}[t]
\includegraphics[width=0.9\linewidth]{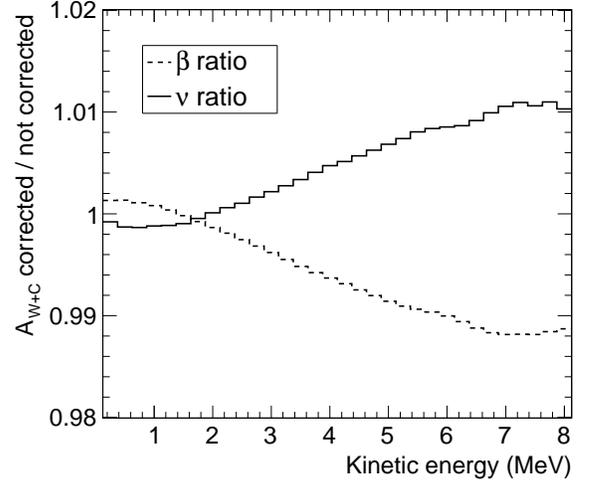}
\caption{Ratio of total spectra computed from all ENSDF branches with and without the $A_C$ and $A_W$ correction terms. 
The dashed (solid) curve shows the ratio of electron (neutrino) spectra.}
\label{fig:AcwRatio}
\end{figure}

\begin{figure}[b]
\includegraphics[width=0.9\linewidth]{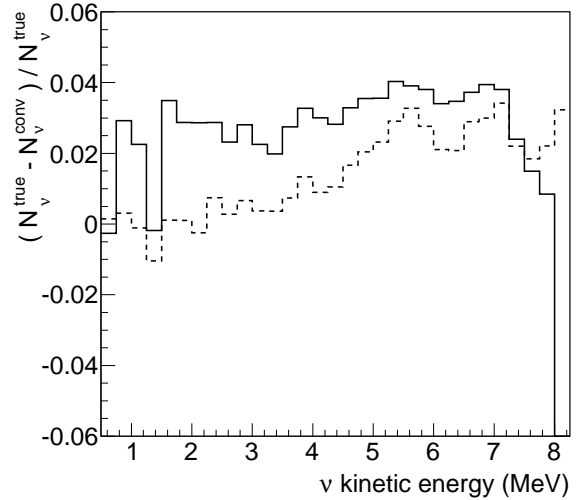}
\caption{The solid curve is the same solid curve as in figure~\ref{fig:TestConv}. The dashed curve illustrates the 
deviation induced by the implementation of $A_C$ and $A_W$ corrections at the level of each virtual branch. It explains 
most of the +3\% shift observed in figures \ref{fig:new_residues}(b) and \ref{fig:TestConv} below 4 MeV.}
\label{fig:AcwEffect}
\end{figure}

\begin{figure}[b]
\includegraphics[width=0.9\linewidth,angle=0]{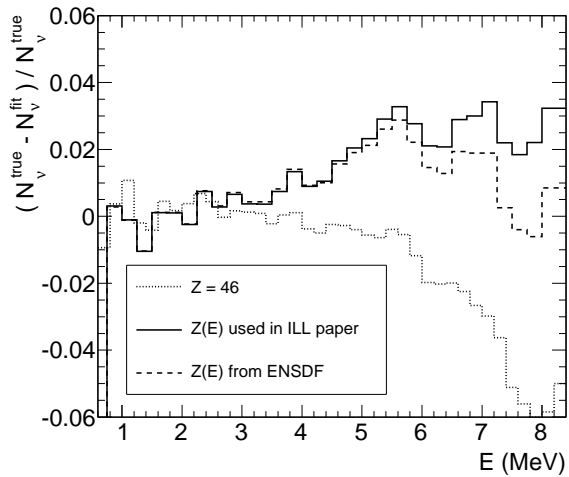}
\caption{Deviation from the true antineutrino spectrum induced by various Z(E) functions used in the formula of the virtual branches. 
For this test $A_C$ and $A_W$ are turned off in the "true" branches as well as in the virtual branches. This effect takes over 
the +3\% deviation observed in figures \ref{fig:new_residues}(b) and \ref{fig:TestConv} above 4 MeV.}
\label{fig:ZofE}
\end{figure}

Switching on and off the various ingredients of the ILL conversion reveals a twofold origin of the normalization shift.
At low energy, the deviation is mainly due to the treatment of the $A_C$ and $A_W$ correction terms. In 
Eq.(\ref{eq:delta}) these two terms appear multiplied by the energy hence one would expect their net contribution to grow 
with energy. Nevertheless $A_C$ itself has some hidden energy dependence via the $Z$ distribution of all branches 
(Eq.\ref{eq:A_C}) while the estimate of $A_W$ is a constant (Eq.\ref{eq:A_W}). To illustrate the size of the total 
correction $A_C+A_W$  we show the ratio of corrected over non-corrected spectra in figure \ref{fig:AcwRatio} in the 
case of total spectra build up from all ENSDF branches. There is a linear trend at high energy but direct comparison 
with the effective linear correction used in previous analysis  Eq.(\ref{eq:eff_delta}) shouldn't be made at this stage. 
In fact in the conversion procedure the electron spectrum is fitted on the electron ILL data. By definition the fitting 
procedure optimizes the parameters of the few virtual electron branches used in this work so that the total electron 
spectrum always matches the ILL data, whatever the correction terms included in the theoretical expression of the branches. 
Therefore in the conversion process only the neutrino spectrum is sensitive to the $A_C$ and $A_W$ terms. The final 
effect is not intuitive but can be computed numerically as shown in figure~\ref{fig:AcwEffect}. When applied at the 
branch level the $A_C$ and $A_W$ corrections deviate from the effective formula of Eq.(\ref{eq:eff_delta}) and explain 
the predicted 3\% shift in the low energy region. We stress that no improvement was brought to the theoretical expression 
of the corrections (Eq. \ref{eq:A_C} and \ref{eq:A_W}), only the way of implementing these corrections has been revisited 
avoiding the extra approximation of the effective correction of Eq.(\ref{eq:eff_delta}). While justified at high energy, 
we show that this approximation was giving too much amplitude to the correction at low energy. The effect is large enough 
to explain the observed 3\% shift at low energy.\\
At higher energy the dominant source of the shift comes from the parameterization of the charge distribution associated 
with the virtual $\beta$-branches. Figure~\ref{fig:ZofE} illustrates how large errors can be induced by the too rough 
approximation of a constant nuclear charge. In the ILL data analysis the mean charge of each virtual branch was taken 
from a polynomial fit $Z(E_0)$ of the tabulated nuclear data (Eq (3) of \cite{SchreckU5-2}). This greatly reduced the bias but 
still this approach doesn't take into account the very large dispersion of nuclear charges around this mean. Even a new 
$Z(E_0)$ function, fitted on the same data used to build our "true" spectrum generates a deviation of a few percent at 
high energy.

\begin{figure}[t]
\includegraphics[width=0.9\linewidth]{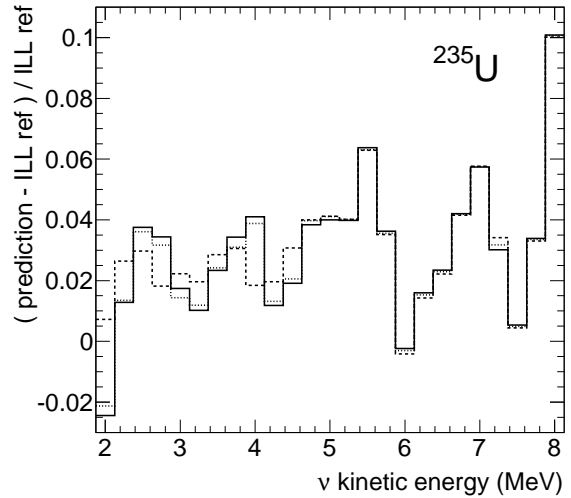}
\caption{Stability check of antineutrino residues when using different estimates of fission yields for the initial {\it ab initio} 
calculation (see eq.~\ref{eq:Sk}): independent yields calculated via MURE after 12h of irradiation (solid line), after 36h 
(dotted line) and cumulative yields (dashed line). The difference between neutrino residues is less than 1\% below 
4.5 MeV and negligible for higher energies}
\label{fig:ingredients}
\end{figure}

\begin{table}[t]
\begin{center}
\medskip
\begin{tabular}{|c|c|c|c|c|c|c|c|}
\hline\hline
$E_{\text{kin}}$  & $\beta$ res. & $N_{\bar\nu_e}$ & \multicolumn{5}{c|}{Error $\Delta N_\nu$ in \% at 1$\sigma$ level} \\
(MeV) & (\%) &  (/fission) &  Stat. & Conv. & $A_{C,W}$ & Norm. & Total \\
\hline
2.00 & ~0.03 & 1.31       & 0.3 & 1.0 & 0.5 & 1.7 & 2.1 \\
2.25 & -0.03 & 1.11       & 0.3 & 1.0 & 0.5 & 1.7 & 2.1 \\
2.50 & ~0.07 & 9.27$(-1)$ & 0.3 & 1.0 & 0.5 & 1.7 & 2.1 \\
2.75 & -0.35 & 7.75$(-1)$ & 0.3 & 1.0 & 0.5 & 1.7 & 2.1 \\
3.00 & ~0.24 & 6.51$(-1)$ & 0.3 & 1.0 & 0.5 & 1.8 & 2.1 \\
3.25 & ~0.14 & 5.47$(-1)$ & 0.3 & 1.0 & 0.5 & 1.8 & 2.1 \\
3.50 & -0.06 & 4.49$(-1)$ & 0.3 & 1.0 & 1.0 & 1.8 & 2.3 \\
3.75 & -0.22 & 3.63$(-1)$ & 0.3 & 1.0 & 1.0 & 1.8 & 2.3 \\
4.00 & -0.19 & 2.88$(-1)$ & 0.3 & 1.0 & 1.0 & 1.8 & 2.3 \\
4.25 & ~0.52 & 2.27$(-1)$ & 0.3 & 1.5 & 1.0 & 1.8 & 2.6 \\
4.50 & ~0.89 & 1.77$(-1)$ & 0.3 & 1.5 & 1.0 & 1.8 & 2.6 \\
4.75 & ~0.46 & 1.37$(-1)$ & 0.3 & 1.5 & 1.0 & 1.8 & 2.6 \\
5.00 & ~0.70 & 1.09$(-1)$ & 0.3 & 1.5 & 1.5 & 1.8 & 2.8 \\
5.25 & ~0.43 & 8.54$(-2)$ & 0.3 & 1.5 & 1.5 & 1.8 & 2.8 \\
5.50 & -1.24 & 6.56$(-2)$ & 0.3 & 3.0 & 1.5 & 1.8 & 3.8 \\
5.75 & -1.56 & 4.99$(-2)$ & 0.3 & 3.0 & 2.0 & 1.8 & 4.1 \\
6.00 & -0.59 & 3.68$(-2)$ & 0.3 & 3.0 & 2.0 & 1.8 & 4.1 \\
6.25 & -0.62 & 2.74$(-2)$ & 0.3 & 3.0 & 2.0 & 1.9 & 4.1 \\
6.50 & -0.08 & 2.07$(-2)$ & 0.3 & 3.0 & 2.0 & 1.9 & 4.1 \\
6.75 & ~0.09 & 1.56$(-2)$ & 0.3 & 3.0 & 2.0 & 1.9 & 4.1 \\
7.00 & -0.27 & 1.11$(-2)$ & 0.4 & 3.0 & 2.0 & 1.9 & 4.1 \\
7.25 & -0.90 & 6.91$(-3)$ & 0.4 & 3.0 & 2.0 & 1.9 & 4.1 \\
7.50 & -0.93 & 4.30$(-3)$ & 0.5 & 3.0 & 2.5 & 1.9 & 4.4 \\
7.75 & -0.14 & 2.78$(-3)$ & 0.9 & 3.0 & 2.5 & 1.9 & 4.4 \\
8.00 & -1.18 & 1.49$(-3)$ & 1.8 & 3.0 & 2.5 & 1.9 & 4.7 \\
\hline\hline
\end{tabular}
\caption{
Results of the new conversion procedure on the $^{235}$U antineutrino spectrum. The electron residues between our
prediction and ILL data are given in percent as an indication of the quality of the fitting procedure. The antineutrino
spectrum is normalized per fission and correspond to the spectrum for a 12h irradiation time. All errors are given in
percent at 1$\sigma$ (68\% CL). \label{tab:final_resultsU5}}
\end{center}
\end{table}

Further cross-checks of our results are based on minimizing the electron residues in different independent ways. First
we checked that the new conversion procedure is not sensitive to the chosen starting point for the {\it ab initio}
calculations. All the results presented above were obtained by adding 5 virtual branches to the spectrum built up from
nuclear databases and using independent yields calculated after 12 hours irradiation time to match as closely as
possible the experimental conditions at ILL. Figure~\ref{fig:ingredients} shows the variation of the neutrino residues
when using independent yields at 36 hours instead of 12 hours or even cumulative fission yields, corresponding to the
equilibrium regime reached after infinite irradiation time. One can see that the variations induced in the antineutrino
spectra are negligible ($\leq 1\%$). This can be understood by the fact that although the different sets of fission
yields do change the shape of the {\it ab initio} spectrum by a few percent, this change is absorbed by the virtual
branches fitting the missing contribution with respect to the ILL electron spectrum. The underlying distributions of
nuclear charges and end-points remain very similar leading to the same final residues. This illustrates how our mixed
approach gets rid of the dominant errors of the {\it ab initio} approach.

In order to avoid the use of virtual branches we also tried minimizing the electron residues by tweaking the input
parameters of the {\it ab initio} calculation, namely, the distributions of branching ratios or end-points or both at
the same time. This technique makes sense only if the tweaking does not disturb too much the physical
distributions. Therefore in the minimization procedure we implemented limitations of typically 15\% for the variation
range of physical parameters. Electron residues of similar quality could be obtained in this way at low energy but
deviations of several percents couldn't be avoided above 5 MeV. As already observed in figure~\ref{fig:new_residues},
any large residues pattern in the electron fit shows up in the antineutrino residues, slightly shifted in kinetic energy
and amplified by a factor of about 3. This behavior is used to estimate our error budget, summarized in
table~\ref{tab:final_resultsU5}. Even in the case of zero residues, we count the statistical error of the reference
$\beta$ spectrum as the minimum error of the converted antineutrino spectrum (column~4). Based on our numerous tests of
fitting methods, the amplified envelope of non statistical electron residues is added in quadrature as an extra
conversion error (column ~5).

In column~6 the error due to the A$_\text{C,W}$ terms is computed by propagating a 100\% relative uncertainty through
the conversion procedure. Finally the normalization error of the ILL reference data is directly propagated as a
normalization error of the converted antineutrino spectrum (column~7). The total error is taken as the quadratic sum of
all previous sources of errors. In the perspective of neutrino oscillation analyses it is mandatory to consider the
correlations between energy bins. The statistical and conversion errors are driven by random processes. Therefore they
do not induce any bin to bin correlation. The normalization error of the ILL data should be treated as fully correlated
over the whole energy range. Regarding the $A_{C,W}$ terms we observe that they are propagated as a linear correction to
the converted antineutrino spectrum above $>4\text{ MeV}$. The uncertainty on the slope coefficient fully correlates all
high energy bins. Below 4 MeV the precise determination of the correlations would require dedicated numerical studies
but in that energy domain the size of the $A_{C,W}$ corrections and their error are small and have negligible impact in
the error budget.

\begin{table}[t]
\begin{center}
\medskip
\begin{tabular}{|c|c|c|c|c|c|c|}
\hline\hline
& \multicolumn{3}{c|}{$^{239}$Pu} & \multicolumn{3}{c|}{$^{241}$Pu} \\
$E_{\text{kin}}$ & $\beta$ res & $N_{\bar\nu_e}$ & $\Delta N_{\bar\nu_e}$ & $\beta$ res & $N_{\bar\nu_e}$ & $\Delta N_{\bar\nu_e}$ \\
(MeV)       & (\%)        & (/fission)      & (\%)                 & (\%)        & (/fission)      & (\%) \\ \hline
2.00 & ~-0.03 & 1.13      & ~2.3 & -0.04 & 1.27      & 2.2 \\
2.25 & ~-0.12 & 9.19$(-1)$& ~2.3 & ~0.10 & 1.07      & 2.2 \\
2.50 & ~~0.14 & 7.28$(-1)$& ~2.4 & -0.11 & 9.06$(-1)$& 2.2 \\
2.75 & ~-0.38 & 6.13$(-1)$& ~2.4 & ~0.00 & 7.63$(-1)$& 2.2 \\
3.00 & ~~0.31 & 5.04$(-1)$& ~2.4 & ~0.24 & 6.39$(-1)$& 2.2 \\
3.25 & ~~0.05 & 4.10$(-1)$& ~2.4 & ~0.98 & 5.31$(-1)$& 2.2 \\
3.50 & ~~0.04 & 3.21$(-1)$& ~2.6 & ~0.65 & 4.33$(-1)$& 2.4 \\
3.75 & ~~1.49 & 2.54$(-1)$& ~2.6 & ~0.49 & 3.51$(-1)$& 2.4 \\
4.00 & ~-0.87 & 2.00$(-1)$& ~2.7 & -0.09 & 2.82$(-1)$& 2.5 \\
4.25 & ~-0.63 & 1.51$(-1)$& ~2.9 & ~0.02 & 2.18$(-1)$& 2.7 \\
4.50 & ~~4.49 & 1.10$(-1)$& ~3.0 & -1.26 & 1.65$(-1)$& 2.8 \\
4.75 & ~~0.21 & 7.97$(-2)$& ~3.0 & -0.80 & 1.22$(-1)$& 2.8 \\
5.00 & ~-2.47 & 6.15$(-2)$& ~3.3 & -0.48 & 9.59$(-2)$& 3.1 \\
5.25 & ~-2.48 & 4.68$(-2)$& ~3.3 & -0.92 & 7.36$(-2)$& 3.1 \\
5.50 & ~-5.41 & 3.50$(-2)$& ~4.4 & -0.93 & 5.52$(-2)$& 4.3 \\
5.75 & ~-0.32 & 2.55$(-2)$& ~4.6 & -0.07 & 4.01$(-2)$& 4.5 \\
6.00 & ~~2.26 & 1.82$(-2)$& ~4.9 & ~1.69 & 2.81$(-2)$& 4.7 \\
6.25 & ~~1.53 & 1.32$(-2)$& ~5.0 & ~0.77 & 2.04$(-2)$& 4.7 \\
6.50 & ~-0.76 & 9.82$(-3)$& ~5.2 & ~0.10 & 1.50$(-2)$& 4.9 \\
6.75 & ~~5.47 & 7.32$(-3)$& ~5.2 & -0.94 & 1.07$(-2)$& 4.9 \\
7.00 & ~-5.01 & 5.13$(-3)$& ~7.1 & -0.32 & 7.20$(-3)$& 5.3 \\
7.25 & ~-1.73 & 3.15$(-3)$& ~9.2 & -1.23 & 4.47$(-3)$& 5.3 \\
7.50 & ~-8.94 & 1.83$(-3)$& 11.1 & -0.96 & 2.54$(-3)$& 5.7 \\
7.75 & -32.75 & 1.03$(-3)$& 15.7 & -1.07 & 1.65$(-3)$& 5.7 \\
8.00 & -55.56 & 4.91$(-4)$& 20.6 & -1.55 & 9.63$(-4)$& 7.0 \\
\hline\hline
\end{tabular}
\caption{
Results of the new conversion procedure on $^{239}$Pu (1.5 days irradiation time) and $^{241}$Pu (1.8 days irradiation
time) antineutrino spectra - see comments for $^{235}$U. The conversion error and the error due to the A$_\text{C,W}$
terms are the same as those given in table~\ref{tab:final_resultsU5}. The statistical and normalization errors for both
plutonium isotopes can be found in~\cite{SchreckPu9Pu1}. \label{tab:resultsPu} }
\end{center}
\end{table}

We applied our same conversion procedure to plutonium isotopes. Results are given in
table~\ref{tab:resultsPu}. The conclusions given for the $^{235}$U antineutrino spectrum remain valid for these
isotopes. The main net effect is a mean $\approx +3\%$ normalization shift with respect to previous reference spectra. 
The equivalent increase in the detected neutrino spectrum is 3.1\% for $^{239}$Pu and 3.7\% for $^{241}$Pu.

\section{Context of reactor neutrino experiments}
\label{sec:ReactorNuExp}

\subsection{A useful phenomenological parameterization}
\label{subsec:Param}

\begin{figure}[!t]
\includegraphics[width=.90\linewidth]{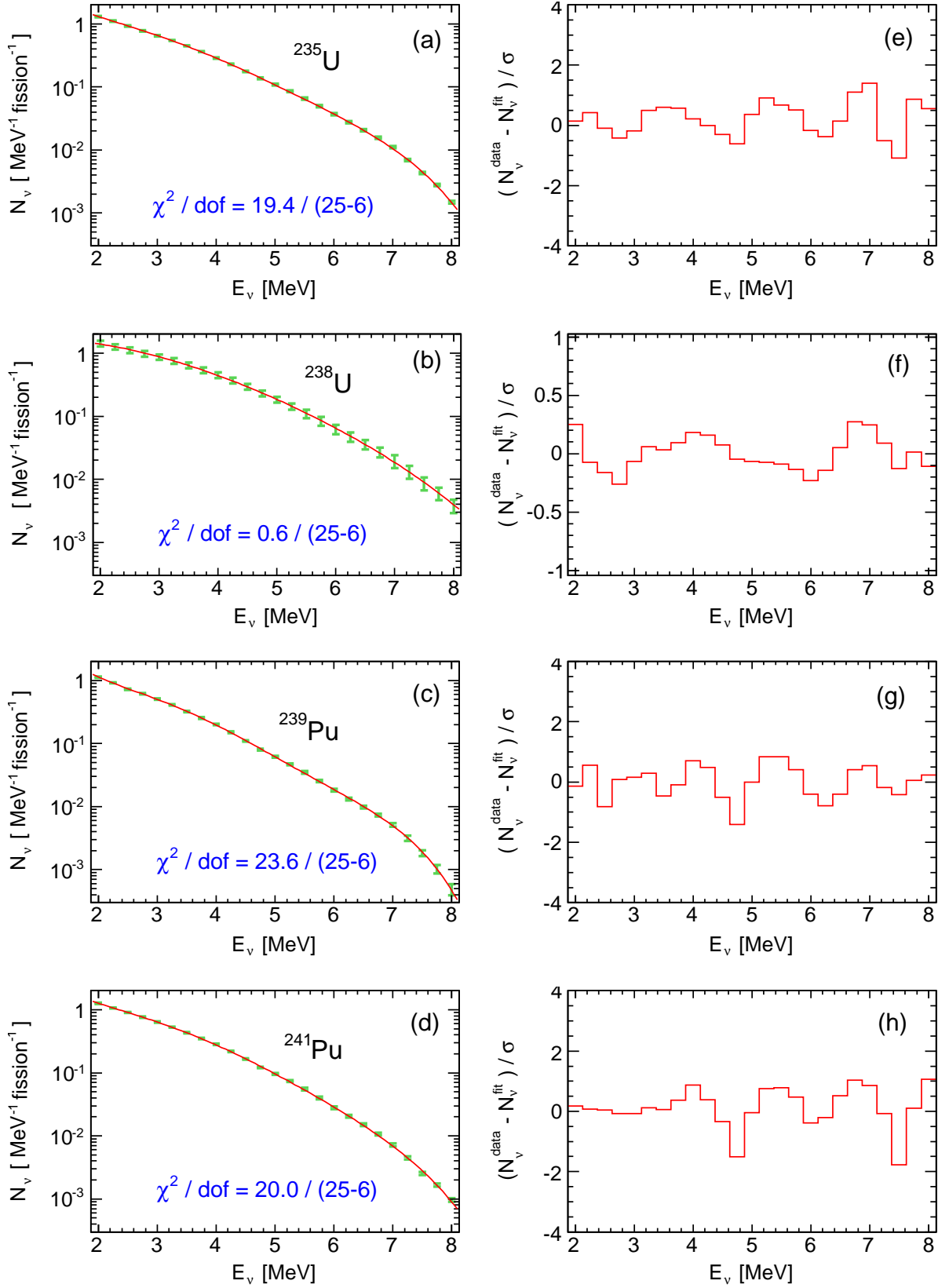}
\caption{(Color on line) Illustration of the fit of the antineutrino spectra predicted in this work. The red curves in the left panels correspond to a 6 parameters fit (polynomial of order 5). Also shown the data with their $1\sigma$ error bars and the $\chi^2$ per degree of freedom. In the right panels we show the residual of the fits.}
\label{fig:fitHetS}
\end{figure}

A phenomenological parameterization of our fission antineutrino spectra could be useful for sensitivity studies requiring
different binning or energy domains than those proposed in tables~\ref{tab:238U_spectra},~\ref{tab:final_resultsU5}
and~\ref{tab:resultsPu}. Therefore, as in~\cite{Huber:2004xh} we provide a parameterization of the spectrum of a given 
isotope using the exponential of a polynomial
\beq
\label{eq:ParamSpec}
S_{k\text{,fit}}(E_\nu) = \exp \Big( \sum_{p=1}^{6} \alpha_{pk}E_\nu^{p-1}\Big).
\eeq
with the coefficients $\alpha_{pk}$ determined by a fit to the data using the MIGRAD algorithm of the TMinuit ROOT class~\cite{minuit}. 
To this aim we minimize the $\chi^2$-function
\beqn
\nonumber \chi^2 & = & \sum_{i,j}D_iV_{ij}^{-1}D_j \\
& & \text{ with } D_i \equiv \sum_{p=1}^{6}\alpha_{pk}(E_\nu^{(i)})^{p-1}-\ln S_k^{(i)}
\eeqn
where $E_\nu^{(i)}$ and $S_k^{(i)}\equiv S_k(E_\nu^{(i)})$ are the values of the antineutrino energy and the corresponding 
antineutrino spectrum, respectively, provided in tables~\ref{tab:238U_spectra},~\ref{tab:final_resultsU5} and~\ref{tab:resultsPu}. 
Since we are fitting the logarithm of the flux the covariance matrix $V_{ij}$ contains the  relative errors of the $S_k^{(i)}$. For the 
diagonal element $V_{ii}$ we take the total error quoted in the above mentioned tables. Because the error of the $^{238}$U spectrum 
has been estimated from the envelope of all systematic effects of the nuclear databases, we assume no bin to bin correlation. For 
our new converted spectra, the fully correlated errors on the absolute calibration of ILL $\beta$ spectra and on the A$_\text{C,W}$ 
correction terms contribute to the off-diagonal elements of the covariance matrix as
\beq
V_{ij}=\sigma_i^\text{cal}\sigma_j^\text{cal}+\sigma_i^\text{corr}\sigma_j^\text{corr}, \qquad i\neq j.
\eeq
The plots (a) to (d) of figure~\ref{fig:fitHetS} shows the resulting spectra for the 6 parameter fits in comparison to the
data and the corresponding $\chi^2$-values per degree of freedom. The best goodness-of-fit is obtained with polynomial
of order five. The plots (e) to (h) of figure~\ref{fig:fitHetS} show the residues of the fit in units of $\sigma_i$, where
the error is obtained from the covariance matrix $V$ by $\sigma_i=S^{(i)}_{k}\sqrt{V_{ii}}$. Note that in case of
correlations between the $S_k^{(i)}$ these residuals do not add up to the total $\chi^2$. All antineutrino spectra are
very well described by the chosen phenomenological parameterization of Eq.(\ref{eq:ParamSpec}). The best fit coefficients
$\alpha_{pk}$ and their correlation matrix are given in table~\ref{tab:fitHetS}. We can see quite large anticorrelations among consecutives 
fit parameters which could be induced by the choice of the exponential of a polynomial for the fit function. Therefore one should be aware 
of possible bias in the propagation of correlations when using these fits whereas it is a practical way to compute nominal spectra.

\begin{table}[h]
\begin{center}
\begin{tabular}{|c|c|c|cccccc|}
\hline\hline
\multicolumn{3}{|c|}{$k$ = $^{235}$U} & \multicolumn{6}{c|}{correlation matrix $\rho^k_{pp^\prime}$} \\
\hline
$p$ & $\alpha_{pk}$      & $\delta\alpha_{pk}$ &  1     &  2     &  3     &  4     &  5     &  6     \\
\hline
  1 &  3.217     & 4.09(-2)    &  1.00 & -0.86 &  0.60 &  0.07 & -0.17 & -0.14 \\
  2 & -3.111     & 2.34(-2)    & -0.86 &  1.00 & -0.84 &  0.12 &  0.25 &  0.01 \\
  3 &  1.395     & 4.88(-3)    &  0.60 & -0.84 &  1.00 & -0.56 & -0.19 &  0.24 \\
  4 & -3.690(-1) & 6.08(-4)    &  0.07 &  0.12 & -0.56 &  1.00 & -0.42 & -0.14 \\
  5 &  4.445(-2) & 7.77(-5)    & -0.17 &  0.25 & -0.19 & -0.42 &  1.00 & -0.77 \\
  6 & -2.053(-3) & 6.79(-6)    & -0.14 &  0.01 &  0.24 & -0.14 & -0.77 &  1.00 \\
\hline\hline
\multicolumn{3}{|c|}{$k$ = $^{238}$U} & \multicolumn{6}{c|}{correlation matrix $\rho^k_{pp^\prime}$} \\
\hline
$p$ & $\alpha_{pk}$      & $\delta\alpha_{pk}$ &  1     &  2     &  3     &  4     &  5     &  6     \\
\hline
  1 &  4.833(-1) & 1.24(-1)    &  1.00 & -0.86 &  0.20 &  0.30 &  0.08 & -0.27 \\
  2 &  1.927(-1) & 5.86(-2)    & -0.86 &  1.00 & -0.58 & -0.21 &  0.04 &  0.23 \\
  3 & -1.283(-1) & 1.11(-2)    &  0.20 & -0.58 &  1.00 & -0.48 & -0.17 &  0.20 \\
  4 & -6.762(-3) & 1.92(-3)    &  0.30 & -0.21 & -0.48 &  1.00 & -0.36 & -0.20 \\
  5 &  2.233(-3) & 2.84(-4)    &  0.08 &  0.04 & -0.17 & -0.36 &  1.00 & -0.77 \\
  6 & -1.536(-4) & 2.86(-5)    & -0.27 &  0.23 &  0.20 & -0.20 & -0.77 &  1.00 \\
\hline\hline
\multicolumn{3}{|c|}{$k$ = $^{239}$Pu} & \multicolumn{6}{c|}{correlation matrix $\rho^k_{pp^\prime}$} \\
\hline
$p$ & $\alpha_{pk}$      & $\delta\alpha_{pk}$ &  1     &  2     &  3     &  4     &  5     &  6      \\
\hline
  1 &  6.413  & 4.57(-2)    &  1.00 & -0.86 &  0.60 &  0.10 & -0.17 & -0.13  \\
  2 & -7.432  & 2.85(-2)    & -0.86 &  1.00 & -0.84 &  0.08 &  0.25 & -0.01  \\
  3 &  3.535  & 6.44(-3)    &  0.60 & -0.84 &  1.00 & -0.54 & -0.20 &  0.26  \\
  4 & -8.820(-1) & 9.11(-4)    &  0.10 &  0.08 & -0.54 &  1.00 & -0.45 & -0.08  \\
  5 &  1.025(-1) & 1.38(-4)    & -0.17 &  0.25 & -0.20 & -0.45 &  1.00 & -0.79  \\
  6 & -4.550(-3) & 1.29(-5)    & -0.13 & -0.01 &  0.26 & -0.08 & -0.79 &  1.00  \\
\hline\hline
\multicolumn{3}{|c|}{$k$ = $^{241}$Pu} & \multicolumn{6}{c|}{correlation matrix $\rho^k_{pp^\prime}$} \\
\hline
$p$ & $\alpha_{pk}$      & $\delta\alpha_{pk}$ &  1     &  2     &  3     &  4     &  5     &  6      \\
\hline
  1 &  3.251  & 4.37(-2)    &  1.00 &  0.87 & -0.60 & -0.08 &  0.17 &  0.13  \\
  2 & -3.204  & 2.60(-2)    &  0.87 &  1.00 & -0.84 &  0.11 &  0.25 & -0.00  \\
  3 &  1.428  & 5.66(-3)    & -0.60 & -0.84 &  1.00 & -0.56 & -0.19 &  0.26  \\
  4 & -3.675(-1) & 7.49(-4)    & -0.08 &  0.11 & -0.56 &  1.00 & -0.43 & -0.11  \\
  5 &  4.254(-2) & 1.02(-4)    &  0.17 &  0.25 & -0.19 & -0.43 &  1.00 & -0.78  \\
  6 & -1.896(-3) & 9.03(-6)    &  0.13 &  0.00 &  0.26 & -0.11 & -0.78 &  1.00  \\
\hline\hline
\end{tabular}
\caption{ Coefficients $\alpha_{pk}$ of the polynomial of order 5 for antineutrino flux from elements $k$ = $^{235}$U, $^{238}$U, 
$^{239}$Pu and $^{241}$Pu. In the column $\delta\alpha_{pk}$ the $1\sigma$ error on $\alpha_{pk}$ are given. Furthermore the 
correlation matrix of the errors is shown.\label{tab:fitHetS}}
\end{center}
\end{table}

\newpage
\onecolumngrid

\clearpage

\twocolumngrid

\subsection{Off-equilibrium corrections}
\label{subsec:off-eq}

The ILL spectra were acquired after a quite short irradiation time in a quasi pure thermal neutron flux, between 12 hours and 1.8 days
depending on the measured isotopes. For neutrino reactor experiments the irradiation time scale would rather be a reactor cycle duration, 
typically 1 year. Among the fission products, about 10\% of them have a $\beta$-decay life-time long enough to keep accumulating after 
several days some of them presenting sufficient large capture cross sections to possibly affect the final inventory. Moreover, in a standard 
PWR, the neutron energy spectrum exhibits more important epithermal and fast neutron energy components than in the ILL measurements. 
These higher energy components of the neutron flux add small contributions to the fissions of $^{235}$U (as well as for the other 
fissioning isotopes) leading to different distributions of the fission products. In this section we study the effect of these phenomena on 
the reference neutrino spectra and compute the associated corrections. Since these corrections are relative deviations between spectra at 
different irradiation times, we assume they are pretty insensitive to the sources of error of our {\it ab initio} calculations discussed in 
section~\ref{sec:micro_comp}.

Therefore the study was done with the MURE simulation of a PWR assembly of N4 type exhibiting a moderation ratio equal to the one of a 
PWR core in order to represent the full reactor neutronic conditions. The infinite multiplication coefficient of the simulation has been 
successfully compared with similar simulations performed with the deterministic code DRAGON for french PWRs \cite{Lemer}. This simulation represents thus very well the real physical conditions of a reactor core. Condition of a constant power is assumed, renormalizing
the neutron flux at each time step in order to compensate for the fuel burnup. We adapted the code in order to compute and 
store the amount of all $\beta^-$ emitters produced over time \cite{ND2007}. In our simulation, the fission yields from the JEFF3.1.1 nuclear
data library \cite{JEFF} were used. The yields coming from the 25~meV, 400~keV and 14~MeV libraries were weighted by the fission rates in 
each neutron energy region. This simulation was compared with independent calculations using the FISPACT code \cite{Fispact} based on the 
EAF nuclear data library which just evolves the isotopic concentrations over time. A constant mean neutron flux of 
$3.10^{14} \text{neutron/cm}^2\text{/s}$ was used for the FISPACT calculations. 

\begin{table}[h]
\centering
\begin{tabular}{|c|c|c|c|c|c|}
\hline
\hline
\multicolumn{6}{|c|}{$^{235}$U} \\   
\hline
  & 2.0 MeV & 2.5 MeV & 3.0 MeV & 3.5 MeV & 4.0 MeV \\ 
\hline
36 h    & 3.1 & 2.2 & 0.8 & 0.6 & 0.1 \\
100 d   & 4.5 & 3.2 & 1.1 & 0.7 & 0.1 \\
1E7 s   & 4.6 & 3.3 & 1.1 & 0.7 & 0.1 \\
300 d   & 5.3 & 4.0 & 1.3 & 0.7 & 0.1 \\
450 d   & 5.7 & 4.4 & 1.5 & 0.7 & 0.1 \\
\hline
\hline
\multicolumn{6}{|c|}{$^{239}$Pu} \\   
\hline
  & 2.0 MeV & 2.5 MeV & 3.0 MeV & 3.5 MeV & 4.0 MeV \\ 
\hline
100 d & 1.2 & 0.7  & 0.2 & $<0.1$  & $<0.1$ \\
1E7 s & 1.3 & 0.7  & 0.2 & $<0.1$  & $<0.1$ \\
300 d & 1.8 & 1.4  & 0.4 & $<0.1$ & $<0.1$ \\
450 d & 2.1 & 1.7  & 0.5 & $<0.1$  & $<0.1$ \\
\hline
\hline
\multicolumn{6}{|c|}{$^{241}$Pu} \\   
\hline
  & 2.0 MeV & 2.5 MeV & 3.0 MeV & 3.5 MeV & 4.0 MeV \\ 
\hline
100 d & 1.0 & 0.5 & 0.2 & $<0.1$ & $<0.1$ \\
1E7 s & 1.0 & 0.6 & 0.3 & $<0.1$ & $<0.1$ \\
300 d & 1.6 & 1.1 & 0.4 & $<0.1$ & $<0.1$ \\
450 d & 1.9 & 1.5 & 0.5 & $<0.1$ & $<0.1$ \\
\hline
\hline
\end{tabular}
\caption{
Relative off-equilibrium correction (in \%) to be applied to the reference antineutrino spectra listed in tables~\ref{tab:final_resultsU5} 
and~\ref{tab:resultsPu}, for several energy bins and several irradiation times significantly longer than the reference times 
(12h U for and 36h for Pu). Effect of neutron captures on fission products are included and computed using the simulation 
of a PWR fuel assembly with the MURE code. \label{tab:off_eq2}}
\end{table}

The departures from our reference spectra are displayed as a function of time in table \ref{tab:off_eq2} for some relevant low energy bins.
As expected, the accumulation of long-lived nuclei shows up as positive deviations which amplitude decreases with the neutrino energy 
and becomes negligible above 3.5 MeV. At the threshold of the beta-inverse process it takes about 100 days of irradiations for the 
antineutrino spectrum to be stable at the 1\% level. Noting that the irradiation time for the reference spectrum of $^{235}\text{U}$ is 12~h 
instead of 36~h for $^{239}\text{Pu}$ and $^{241}\text{Pu}$, the corrections are similar for all isotopes. We checked with our evolution 
codes that the effects of neutron capture on the fission products as well as the contribution of the neutron spectrum above the thermal 
energy domain have small impact on the off-equilibrium corrections. Other tests have been performed with the FISPACT code, showing 
that these results may depend on the system (neutron, flux and energy spectrum, geometry) used in the calculation. The error envelop 
covering our different results is of 30\% on the total off-equilibrium corrections. Therefore the results quoted in table~\ref{tab:off_eq2} 
should be taken as typical corrections at a N4 reactor. For applications with signicantly different neutron flux or fuel geometry, dedicated 
simulations should be carried out for an accurate correction of the lowest energy bins of the antineutrino spectrum.

\begin{figure}[h]
\includegraphics[width=0.9\linewidth]{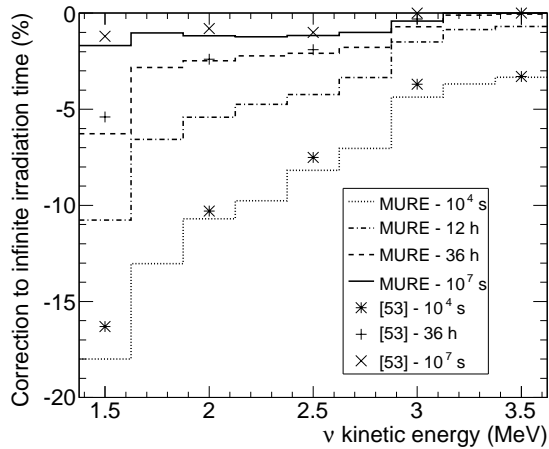}
\caption{
Variations of the $^{235}$U antineutrino spectrum for different irradiation times with respect to a reference spectrum considered at equilibrium.}
\label{fig:OffEqEffects}
\end{figure}

Off-equilibrium effects have independently been evaluated for the analysis of the Chooz experiment~\cite{Chooz}, which measured the neutrino 
spectrum of the two N4 reactors of the Chooz site. In this reference, the departure from the antineutrino ILL spectra were computed using 
the cumulative yields of some known long-lived fission fragments. The results are shown as markers in figure~\ref{fig:OffEqEffects} to be 
compared with the histograms of our calculations. The overall agreement is good, even when evolving the spectrum back to irradiation time as 
short as $10^4$~s, where the corrections become quite large and have steep variations in time. 

Note that our reference spectrum at equilibrium does not use cumulative yields. Instead it is computed by our evolution code using independent 
fission yields and a long irradiation time. In the analysis of cumulative yields it is in fact assumed that all nuclei have reached equilibrium. This is 
fully justified for short lived fission products, whereas there is some ambiguity with the decay products of longest half-lives. To avoid apparent 
double counting of the cumulative yields of the daughters, some very long-lived decays have been removed from the cumulative yield databases 
but some are still present in the libraries. Because of these problems, it is recommended by nuclear databases~\cite{tecdocIAEA1168} as a safer 
method to use independent yields with an inventory code. We have considered 450 days of irradiation as our reference to account for a spectrum 
that would have reached quasi-equilibrium. In addition, this duration corresponds to a typical irradiation time of a fuel assembly in a PWR core. 

Off-equilibrium effects where also computed in~\cite{KopeikinArXiv} where the authors used beta branches of 571 fission fragments. The fission 
yield were taken from~\cite{EnglandRider} and beta decay properties came from experimental data. Our results are compatible with theirs considering 
the quoted uncertainties and the possible discrepancies in the neutron energy spectrum and flux used in the calculations. Note also that small 
additional discrepancies could arise from the smaller number of fission products used in the calculation of Kopeikin et~al. \\  

In conclusion, our new reference spectra presented in section \ref{sec:conversion} are, strictly speaking, valid only for irradiation times comparable 
to the ones used at ILL for their measurement. For longer irradiations, corrections to these spectra are listed in table \ref{tab:off_eq2}. The above 
comparison between independent estimates suggest that the systematic errors associated to these corrections are at the sub-percent level relative to 
the total antineutrino spectrum.

\subsection{Impact on published measurements}
\label{subsec:Impact}

The new conversion of the ILL data described in section~\ref{sec:conversion} leads to a 2.5\% increase of the detected antineutrino flux while the 
predicted shape is basically unchanged. The impact of this correction on the analysis of published measurements of antineutrino oscillations at 
reactor is discussed by the authors in a separate article~\cite{ReAna}. The sensitivity of forthcoming reactor experiments is also updated in this 
context.

\section{Conclusion}
\label{sec:conc}

Using all available data on fission yields and beta decays of fission products we have shown that {\it ab initio} calculations of
total beta spectra agreed with the reference ILL beta spectra at the 10~\% level, illustrating the tremendous amount
of nuclear data collected today. From this work we gave a prediction of the antineutrino spectrum associated with the fission of
$^{238}$U with estimated relative uncertainty increasing from 10 to 20\% with energy in the 2-8 MeV range. Since this isotope
contributes to about 10\% of the total fission rate of a reactor such a prediction is valuable. However, for the dominant isotopes,
the remaining systematic errors of nuclear databases as well as the contribution of poorly known beta transitions still prevent the
{\it ab initio} approach from any use for high precision neutrino oscillation experiments at reactors.\\ This motivated the
development of a new mixed-approach combining the accurate reference of the ILL electron spectra with the physical distribution of
beta branches provided by the nuclear databases. We presented how this method gets rid of the main systematic error of the {\it ab
initio} approach allowing a better control of the conversion of reference electron spectra into antineutrino spectra. While the final
error budget ended up being comparable to previous reference work \cite{SchreckU5-2,SchreckU5Pu9,SchreckPu9Pu1}, we 
demonstrated that the antineutrino spectra emitted by the fission of $^{235}$U, $^{239}$Pu and $^{241}$Pu isotopes have to be 
corrected for a systematic shift of about 3\% in normalization. This net effect was presented as the combination of an improved implementation of finite size corrections to the Fermi theory plus a more realistic description of the distribution of 
$\beta$-branches.\\

\section{Acknowledgments}
We are grateful to M. Cribier for instigating this work. M. Fallot would like to thank J. Wilson and O. M\'eplan for their availibility and 
many useful discussions about the MURE code.

{}

\end{document}